\documentclass[11pt]{article}

\usepackage{geometry}
\usepackage{currfile}
\usepackage{textcomp}
\usepackage{url}

\usepackage{xcolor,soul}
\sethlcolor{yellow}

\usepackage[pdftex]{graphicx}
\DeclareGraphicsExtensions{.pdf,.jpeg,.png}
\graphicspath{{./figures/}}

\date{May 13, 2020}

\begin{document}

\title{\LARGE{An Attestation Architecture for Blockchain Networks\\
\large{ (Extended Abstract) }      }\\
~~}
\author{
\large{Thomas Hardjono$^1$ and Ned Smith$^2$}\\
\large{~~}\\
\large{$^1$Massachusetts Institute of Technology}\\
\large{Cambridge, MA, USA}\\
\small{\tt hardjono@mit.edu}\\
\large{~~}\\
\large{$^2$Intel Corporation}\\
\large{Hillsboro, OR, USA}\\
\small{\tt ned.smith@intel.com}\\
\large{~~}\\
}

\maketitle

\begin{abstract}
If blockchain networks are to become the building blocks 
of the infrastructure for the future digital economy,
then several challenges related to the 
resiliency and survivability of blockchain networks
need to be addressed.
The survivability of a blockchain network is
influenced by the diversity of its nodes.
Trustworthy device-level {\em attestations}
permits nodes in a blockchain network to provide
truthful evidence regarding their current configuration, operational state,
keying material and other system attributes.
In the current work we review the recent developments towards  
a standard attestation architecture and evidence conveyance protocols.
We explore
the applicability and benefits of a standard attestation architecture to blockchain networks.
Finally, we discuss a number of open challenges related to node attestations 
that has arisen due to changing model of blockchain network deployments,
such as the use virtualization and containerization
technologies for nodes in cloud infrastructures.
~~\\
~~\\
Keywords: blockchains, trusted computing, attestations, virtual assets.
\end{abstract}

\newpage
\clearpage

{\small 
\tableofcontents
}

\newpage
\clearpage

\newpage
\clearpage


\section{Introduction}

We believe there is a crucial role for trusted computing technologies,
and more specially attestations technologies,
within the nascent area of blockchain networks.
As blockchain networks play an increasing role
in the future digital economy~\cite{Pentland2020a}
-- such as becoming the underlying infrastructure for future crypto-currencies
and virtual assets exchange networks --
the security, resiliency and survivability
of blockchain systems becomes crucial to their business value-proposition.
Since the dawn of the computer age and the development
of networked computer systems and the Internet,
there has been the need for operators of computing equipment
to obtain correct and truthful insights into the state of computing devices
as part of managing the security of these devices.
Given the proliferation of malware and viruses in the past decade,
there has been a need for networked devices
to have the capability to report its configuration,
internal state and other parameters in a truthful and unforgeable manner.
The technical term used to describe this process is {\em attestations}.

The goal of the current work is threefold.
The first is to review the current development towards  
a standard attestation architecture in the computer and network industry.
Secondly, to explore
the applicability and benefits of the attestations architecture to nodes in a blockchain network.
Thirdly, we discuss some of the current challenges in attestations 
that has arisen due to changing model of blockchain networks,
such as the use virtualization technologies for nodes in cloud infrastructures.

The subject of attestations is at least
two decades old -- stemming from the industry efforts
around the Trusted Platform Module (TPM hardware)~\cite{TPM2003Design} --
and numerous research papers have been
devoted to this subject.
Because the area of trusted computing has been heavily influenced
by the design of the TPM hardware,
much of the discourse in the broader research literature 
has been focused on one feature or another of the TPM hardware
(e.g. the functions of its PCR registers,
its identity keys,
the Quote protocol,
sealing,
the agility of ciphers, 
and so on).

In the current work instead of providing a TPM-centric technical discussion
around attestations and the required infrastructure to support TPM-based attestations,
our goal is to discuss the notion of attestations
in an accessible and meaningful manner.
As such, we will strive to abstract-up from the various design features of the TPM 
and focus on the intent of some of these features, 
narrowing our interest on those features that support attestation
and its potential use in blockchain networks.
We direct readers to the excellent works of~\cite{Proudler2002,ChallenerYoder2008,Proudler2014}
for a deeper treatment of the TPM and its features.

\section{Attestations of Blockchain Nodes: Motivations}
\label{sec:AttestationsMotivations}

There are a number of possible benefits derived from the use of device-level attestations
in the context of blockchain networks generally.
The ability for a node to provide a truthful, complete and unforgeable report
regarding its configuration, computational state, keying material
and other system attributes provides a foundation to building
trust (technical trust) in the network as a whole.
\begin{itemize}

\item	{\em Node device identification}: 
In some blockchain deployments
(e.g. permissioned blockchain networks)
the ability to identify a node, authenticate and obtained
signed assertions (e.g. reports) from the node provides 
a crucial feature for the manageability of the network.

Attestation evidence must be source-authentic from the device.
This means that attestation evidence or assertions
must be signed by a private-key that is {\em bound to the hardware}
(or a private-key that is derived from a hardware-bound key).
A key that is hardware-bound means that it
cannot be removed from the hardware (i.e. removal attack is uneconomical).
Furthermore,
a hardware-bound key may even be inaccessible (invisible)
to the user's application.
A user's application would instead use keys derived from this hardware-bound key.
An example of this is the certified-keys in the TPM hardware
(e.g. AIK-certified keys)~\cite{HardjonoTPM2008}.

\item	{\em Node device configuration reporting}:
The ability for nodes in a blockchain network
to truthfully report its device configuration
(i.e. hardware, firmware, software)
allows the questions related to fairness in the distribution of computing power
(e.g. hashing rate) and other computing capabilities 
to begin to be addressed~\cite{GervaisKarame2014,EyalSirer2014}.

Detailed reporting of hardware configurations can be achieved
using composite attestations approach (see Section~\ref{subsec:CompositeAttestations}).
For example, the ability for a node to provide truthful and unforgeable evidence regarding
the number of GPU cards (for hash-power increase) based on composite attestations
allows a third-party verifier to glean as to the actual hash-power  available at that node.
This in turn allows a community of node-owners in a permissioned blockchain network,
for example,
to obtain accurate estimations regarding the hash-powers available at each of the members.
Such estimations allows fairness to be more readily achieved.

\item	{\em Diversity of nodes and survivability of blockchain networks}:
The ability for nodes
to truthfully generate unforgeable evidence of its device configuration
permits the question of network {\em diversity} 
-- and therefore network {\em survivability}~\cite{Hardjono2018-IEEEGlobalSummit,HardjonoLipton2020b} --
to begin to be addressed.

The question of the diversity of nodes (i.e. diversity of software stack and hardware)
was also touched upon by NIST in their report (NISTIR~8202, Section~8):
... ``A blockchain network is only as strong as the aggregate of 
all the existing nodes participating in the network. 
If all the nodes share similar hardware, software, geographic location, 
and messaging schema then there exists a certain amount of 
risk associated with the possibility of undiscovered security vulnerabilities''~\cite{NIST-80202-2018}.
If there is one lesson learned from the past two decades
of viruses in PC computers, it is that a relatively homogeneous
network of systems (e.g. Windows only) is less resilient than one 
consisting of a diverse set of operating systems
(e.g. mix of Windows, Linux, MacOS, AIX, etc).

As a blockchain networks carry increasingly valuable
virtual assets~\cite{FATF-Recommendation15-2018},
the question of blockchain resiliency and survivability
becomes crucial to the business value proposition of the blockchain network.

\item	{\em Consensus protocol input}:
The state of a node/device can be used as an additional input parameter
into the consensus-making protocol used in the blockchain network.
The idea is that the node that mines or forges
new blocks shoule be in a compliant or ``healthy'' state.
Here ``compliance'' means that the node is deemed satisfactory
as evaluated against the appraisal policies driving the network.
Thus, for example,
the compliance status of a node could be a factor in (input parameter into) 
the consensus-algorithm of a given blockchain network.

Using the specific example of 
the Proof-of-Stake (PoS) protocol in Ethereum~\cite{Ethereum-POS-2019},
the compliance status of nodes could be an additional factor
in selecting the the PoS validator node
-- in addition to the current selection factors
(e.g. staking-age, randomization, node's stake amount, lowest hash value, etc.).

\item	{\em Confidence in Smart Contract Platforms}:
Smart contracts have been a major feature of attraction
for blockchain platforms such as Ethereum.
However, smart contracts themselves may introduce various unforeseen
weaknesses and vulnerabilities~\cite{AtzeiBartoletti2016}.

For example, one concern pertains to the accurate execution and the correct reporting of the outcome
of the smart contract (at the application level).
One key issue here is that even if a smart contract (visibly readable on a node) 
was digitally signed by its author at the application level,
there is no guarantee when the contract binary
was being loaded into memory to be executed (e.g. by the CPU of the node) 
that the code has not been modified
or contaminated by malware.
Here, the availability of Trusted Execution Environments (TEE) 
capability within the node's hardware
can mitigate this problem to a large extent (e.g. Intel's SGX~\cite{McKeen2016,Costan2017}).
However, attestation evidence must be conveyed by 
the node that: (i) the node possesses true TEE functionality,
and (ii) that the contract code was indeed executed by (inside of) the TEE.

\item	{\em Legal trust framework for operating rules}:
The technical means to provide node attestations can play
an important role in certain types of blockchain networks.
For example,
in a consortium-based permissioned (private) blockchain network,
attestations can be a key factor in the consortium's {\em Legal Trust Framework} (LTF)
that governs the operating rules of the membership.

Operational standards (profiles) that are co-developed and defined by the members
of the consortium -- based on stable technical standards
published by standards-organization such as the TCG, IEEE and IETF --
becomes a quantifiable input into the legal contracts
that make-up the LTF framework of the consortium.
The legal contracts can be specific regarding member's obligations
in the realm of deployment of nodes and related services.
(e.g. members' nodes must use valid {X.509} device certificates at all times).
This co-development of operational standards means easier acceptance by the membership.
Furthermore, a joint development of operational standards
allows costs-sharing among members.
Over time it lowers the shared operational costs of running the network as a whole.

However, if a member is able to misrepresent their contribution 
to the shared operational costs,
then the consortium looses its unbiased governance abilities. 
Attestations among consortium member nodes ensures 
operating rules are applied fairly across all members.
Node attestations essentially provide the basis for {\em technical trust},
which in turn allows {\em business trust} to be attained
by virtue the operational specifications being agreed-upon
and observed in the consortium.
Other related contracts and Service Level Agreements (SLAs)
for the network as a whole can also defined in terms
of these operational capabilities.

\end{itemize}

\section{Overview of the Concept of Attestations}

In this section we briefly review at a high-level
a number concepts which underlie
the notion of device attestations that is core to the area of trusted computing.

\subsection{The Notion of the TCB}

As computer systems evolved in the 1980s and 1990s, and as Local Area Networks (LANs) 
and peripheral devices proliferated 
the question of the security of computers systems became crucial 
in the networked world.  
This is true not only in the context of government and defense sectors,
but also in the broader networked computing world. 
In the mid-1980s efforts such as
Project Athena at MIT~\cite{Rosenstein-MIT-Athena-1988,SteinerNeuman1988}
represented the leading edge of networked computing technology,
and numerous security challenges -- such as scalable authentication and authorization --
were identified in these early years.

In the context of trustworthy computing,
a landmark event in December 1985 was the publication of 
the Trusted Computer System Evaluation Criteria (TCSEC)
by the {U.S.} Department of Defense.
The TCSEC was a significant step forward
because it defined the notion of the {\em Trusted Computing Base} (TCB).
The core notion of the TCB is that if protecting the entire computer systems was
too costly or technically infeasible,
then a portion of the system needs to be isolated
that provides trustworthy behavior.
That is, a domain ``boundary'' needs to be identified or defined in the system
within which security can be guaranteed.
This TCB domain boundary must demarcate ``the security-relevant portions'' of the system.
This concept of the TCB became fundamental to ensuing efforts in the area of trustworthy computing,
and the TCB portion became the focus of attention of new technical innovations
in the following two decades.
All subsequent expressions of trustworthy computing and 
security policy would be described in terms of impact and relevance to the TCB.

Although the TCSEC criteria focused mostly on defining the operating system security domain,
it is important to remember that
the operating system is not the sole TCB component in a computing platform. 
The hardware also plays a significant role, 
notably in the context of memory page isolation where a central tenet is
the idea of kernel-mode isolation (namely {\em ring-0}) 
and application-mode (namely {\em ring-3}) process separation contexts~\cite{Saltzer1974}. 
From the operating system perspective the hardware is thought to be
trusted because the operating system has no alternative way to test and verify that the 
hardware is behaving correctly.

The threat of hardware vulnerability motivated the computing industry 
to form the Trusted Computing Group (TCG)~\cite{TCG-website} in the late 1990s
where the notion of a hardware root-of-trust was used to 
distinguish the security relevant portions of a hardware platform. 
The TCG defined trusted computing 
more organically by building upon 
granular components that were 
described as {\em shielded locations} and {\em protected capabilities}. 
Shielded locations are ``...A place (memory, register, etc.) where it is safe 
to operate on sensitive data; data locations that can be accessed only by protected capabilities''. 
Protected capabilities are ``...the set of commands with exclusive permission to access shielded locations.'' 
By extension, all components that could be classified 
as shielded locations or protected capabilities is 
what defines the hardware security domain wherein the TCB software executes.

\subsection{The Attester and Attesting Environment}

\begin{figure}[t]
\centering
\includegraphics[width=1.0\textwidth, trim={0.0cm 0.0cm 0.0cm 0.0cm}, clip]{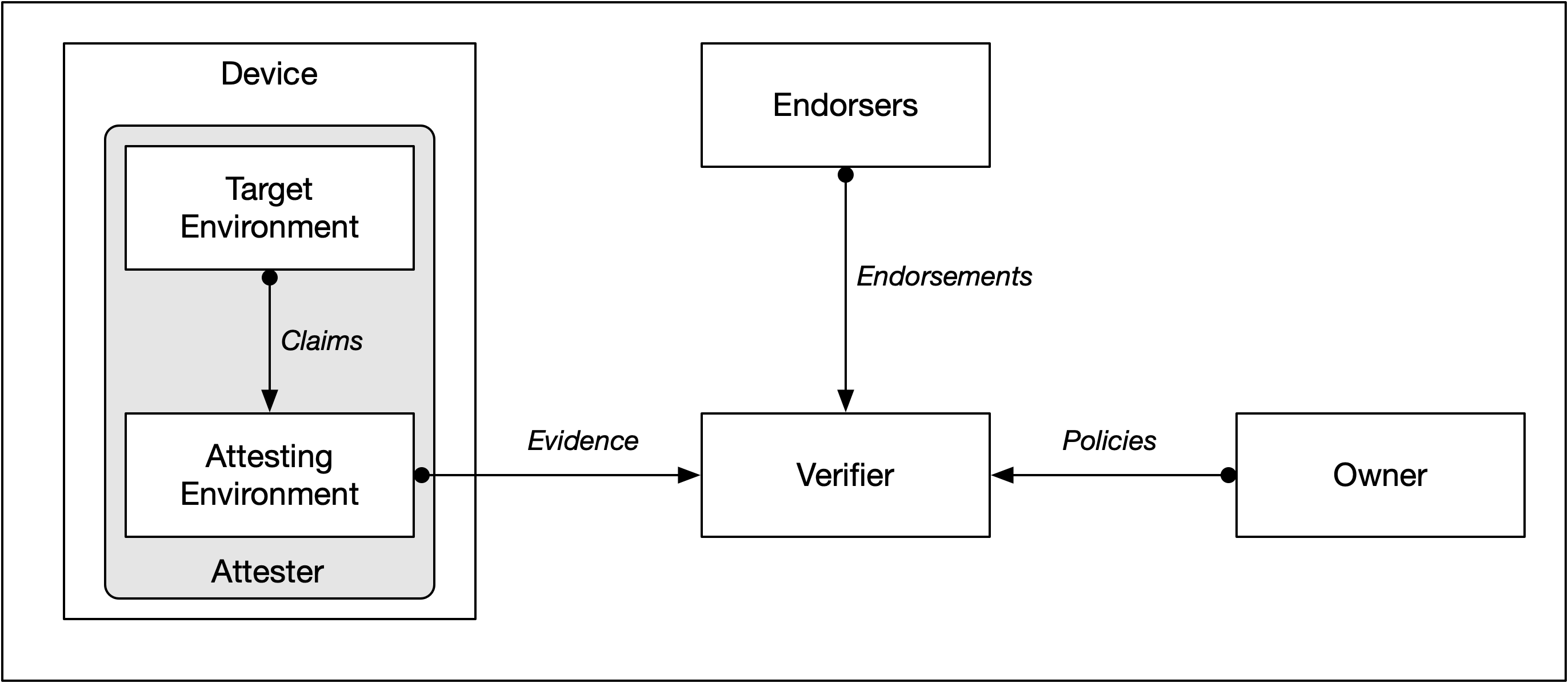}
\caption{Overview of the concept of the Attester and Attesting Environment}
\label{fig:AttestingEnvironment}
\end{figure}

Implied in the TCSEC definition of the TCB is the required ability 
for a TCB to be reviewed. 
Originally, security review was a manual process 
involving a certification process. 
However, subsequent evolution and automation of security review 
makes feasible the reporting of much of the TCBs internal state
-- or what we refer to as {\em attestations}.
The fundamental idea of attestations of a ``thing'' (e.g. a computing device)
is that of the conveyance of truthful information regarding 
the (internal) state of the thing being attested to.
In the related literature on trustworthy computing 
the term ``measurement'' is used to mean the 
act of collecting (introspecting) claims or assertions 
about the internal state, 
and delivering these claims as evidence to an 
external party or entity for automated review and security assessment.

However, as we know today
computing environments can be structurally complex
and may consist of multiple elements (e.g. memory, CPU, storage, networking, firmware, software),
and computational elements can be linked and composed to form computational pipelines, arrays and networks.  
Thus, the dilemma is that not every computational element can be expected to be capable of attestation.
Furthermore, attestation-capable elements may not be capable of attesting every 
computing element with which it interacts.  
The attestation capability could in fact be a computing environment itself.
The act of monitoring trustworthiness attributes, 
collecting them into an interoperable format, 
integrity protecting, authenticating and conveying 
them requires a computing environment -- one that 
should be separate from the one being attested. 
Figure~\ref{fig:AttestingEnvironment} illustrates the recognition of this distinction,
namely of the {\em target environment} being attested to,
and the {\em attesting environment} that performs 
the work stated above\footnote{An example of an attesting environment is
the Quoting Enclave within Intel SGX~\cite{McKeen2013,McKeen2016}.}.

The complexity of the problem has led to a number of efforts in industry
to define an {\em attestation architecture} that incorporates
some of these key concepts -- such as the concept of the root-of-trust --
and to develop standards that implement attestation concepts.
The roles and functions of the attestation architecture is shown in Figure~\ref{fig:AttestingEnvironment}
and will be discussed at length in the ensuing sections.
In a nutshell,
in Figure~\ref{fig:AttestingEnvironment} an attester
conveys evidence of trustworthiness (of the attested target environment)
to a verifier entity.
The verifier operates based on policies that are supplied by the owner of the verifier.

We believe that an attestation architecture should define 
the attestation roles in the ecosystem 
(i.e. Attester, Verifier, Endorser, Relying Party, and Owner), 
the messages they exchange, their structure and the various
ways in  which  roles  may  be hosted, combined and divided amongst 
the various entities involved in real-world deployments. 
These roles should remain true independent of the specific 
use-cases or deployments of systems having attestation functions. 
Furthermore, the attestation messages should be built on 
an information model that defines its trustworthiness semantics 
as well as on a data model that supports broad interoperability options. 
The information model and various data model representations would then 
be realized as data structures, data structure encodings and protocol bindings 
for conveying attestation messages, 
that are aimed at specific deployment cases or ``profiles'' 
(e.g.  PC client devices, constrained IoT devices, 
various network equipment such as routers, mobile devices, server chassis, etc.).

Finally, from architecture perspective we believe 
that the verifier should be able to understand 
the trustworthiness properties of both the target environments 
as well as the attestation capability itself (at the attester). 
This must be true for any set of assertions within 
an attestation flow between the attester and the verifier. 
Thus, the trustworthiness of the attestation capability 
itself should be a core consideration of a well-designed 
attestation architecture. 
This may mean that the attestation architecture 
should anticipate the possibility of 
recursive or layered TCBs,
each having believable and verifiable 
trust properties, something that may complicate implementations considerably.

\subsection{Reference Values and Endorsements}

Another key concept in trustworthy computing is that of {\em endorsements}
from supply-chain entities regarding one or more components
that are incorporated into the target environment.
In practice,
there is a point at which ultimately 
a portion of the computing environment trustworthiness must be established via non-automated means. 
A {\em root-of-trust} refers to a TCB element that ascribes trust through non-automated means. 
These non-automated means include things such
design reviews, manufacturing process audits and physical security.  
A trustworthy attestation mechanism depends on trustworthy manufacturing and supply-chain practices.
This manufacturer's claims of trustworthiness (of its product)
is expressed through endorsements,
which most commonly take the form of digital certificates signed by 
the manufacturer~\cite{TCG-IWG-2005-Thomas-Ned-Editors-part1,TCG-IWG-2006-Thomas-Ned-Editors-part2}.

Thus, a manufacturer of a component (e.g. firmware for hardware component) 
can publish a ``known good'' or ``expected'' value for a given firmware file. 
In the simplest form, this endorsement could be a cryptographic digest 
(i.e. hash) of the firmware file which is authenticated to its 
vendor or place of origin using a digitally signed structure 
(e.g. X.509 attribute certificates~\cite{RFC3281-formatted}).  
The significance of signing by the manufacturer 
using its public-private key pair is that it asserts
these values to be authentic, as an {\em endorsement} for that product. 
Entities seeking to use the product (i.e. firmware file) 
can validate that attested values corresponding to
the digests found in the endorsed values are known or expected. 
Thus, we refer to the digest as a {\em known good value} in this context.

As we will discuss below,
an attestation architecture should distinguish between these more static
endorsements issued by a supply-chain entity
from the {\em evidence} issued by an attester during its runtime.
As shown in Figure~\ref{fig:AttestingEnvironment}
the Attester creates attestation evidence (signed assertions) that are conveyed to a Verifier for appraisal.  
The appraisal process compares the received evidence against the known good values (i.e. endorsements) 
obtained from supply-chain entities -- referred to as {\em endorsers}.  
A good architecture should support multiple forms of appraisals
(e.g. software integrity verification, 
device composition and configuration verification, 
device identity and provenance verification, etc.).  
Out of this appraisal process the attestation {\em results} are generated, 
signed and then conveyed to relying parties.  
The attestation results provide the operational integrity 
basis by which relying parties may determine the level of confidence 
to place in the application data or other application-specific operations that follow.

\section{A Canonical Attestations Architecture}
\label{sec:TCGAttestation}

Recently, the notion of attestations has have garnered interest within
different technical standards organizations and industry consortiums,
beyond the TCG alliance (e.g. FIDO Alliance~\cite{FIDO-key-Attestation-2015}, 
Global-Platform~\cite{GlobalPlatform2012}, IETF~\cite{IETF-RATSWG}).

A broader set of use-cases are also emerging, 
ranging from attestations by routing fabrics to attestations 
by low-power Internet-of-things (IoT) networks. 
Regardless of the use-case, the notion of attestations 
holds true due to the universal need in the digital world 
to obtain assurance regarding the expected working environments 
and their security and resiliency properties. 
As an increasing portion of the economy moves onto digital platforms 
operating using complex and often invisible infrastructures 
(e.g. cloud platforms, virtualization, edge computing, mobile wallets, etc.), 
the more crucial the need for attestation underpinnings 
as a supporting infrastructure.

In this section,
we discuss further the notion of attestations and present a canonical architecture
that is currently being developed by several industry organizations
(e.g. TCG~\cite{TCG-ATTEST2020}, IETF~\cite{IETF-RATSWG}).
The hope is that a canonical attestation architecture 
will allow standards to be developed that implement
the various protocols and flows  
for relevant sectors and products
(routers and network equipment~\cite{TCG-RIV-2020,IETF-rats-network-device-attestation-05},
mobile devices~\cite{GlobalPlatform2012}, 
cloud stacks~\cite{OpenCompute-website2020}, etc).
By having a common reference architecture,
different efforts can share common terminologies, concepts and implementations
and therefore affect a reduction in costs of developing and deploying
the infrastructures supporting cyber-resilience and trustworthy computing generally.

\begin{figure}[t]
\centering
\includegraphics[width=1.0\textwidth, trim={0.0cm 0.0cm 0.0cm 0.0cm}, clip]{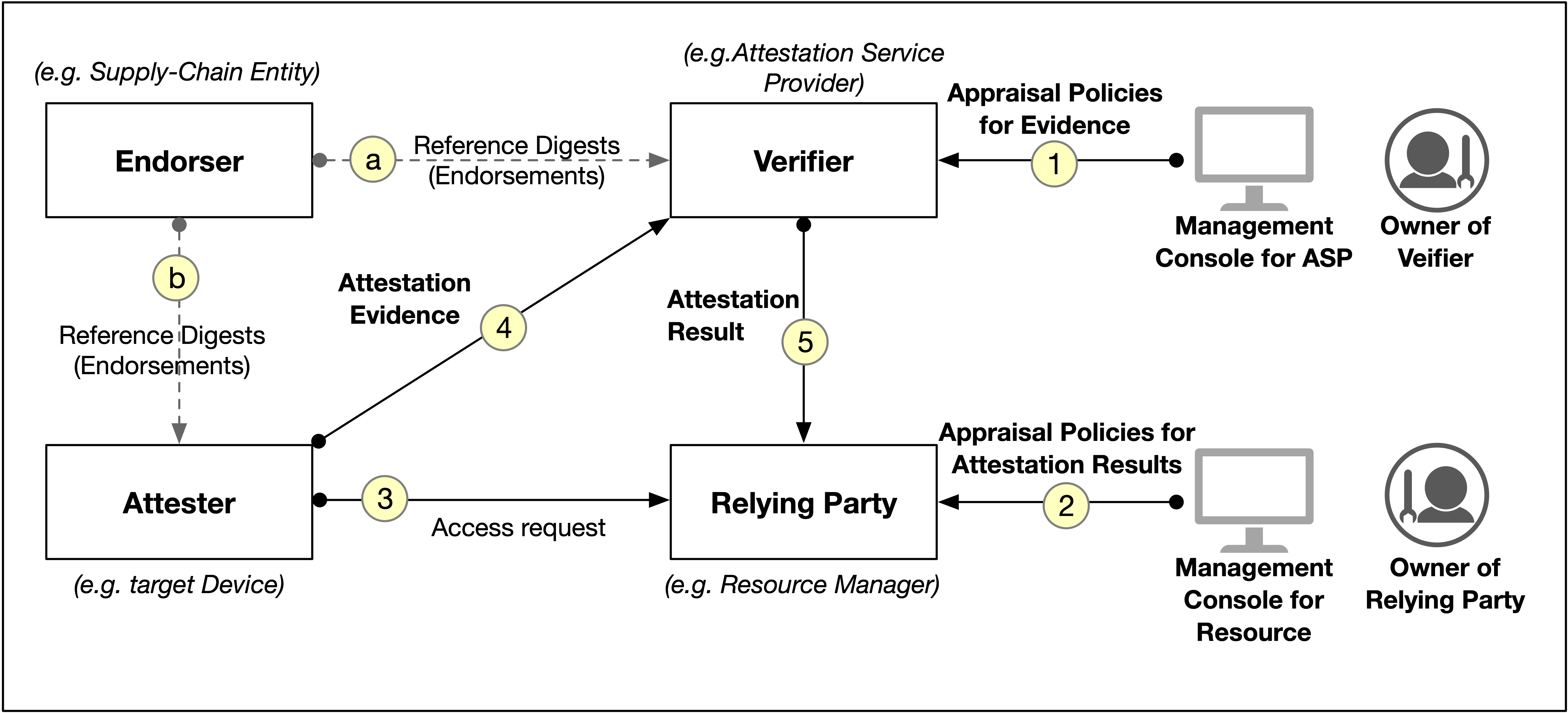}
\caption{Canonical architecture for attestations (after~\cite{TCG-Attestations-Arch2020,rats-arch-02,CokerGuttman2011})}
\label{fig:TCG-Canonical-Model}
\end{figure}

\subsection{Entities, Roles and Actors}
\label{subsec:TCG-Attestation-entities}

The attestation architecture of~\cite{TCG-Attestations-Arch2020}
defines of a set of {\em roles} that 
implement attestation flows.  Roles are hosted by {\em actors},
where actors are deployment entities.  
Different deployment models may coalesce or separate various actor 
components and may call for differing attestation conveyance mechanisms.  
However, different deployment models do not fundamentally modify attestation roles, 
the responsibilities of each role, nor the information that flows between them.
In the following sections, we may use the actor and role terminology interchangeably 
when appropriate in order to simplify discussion (see Figure~\ref{fig:TCG-Canonical-Model}).
\begin{itemize}

\item	{\em Attester}: 
The Attester (e.g. target device) provides attestation Evidence to a Verifier. 
The Attester must have an attestation identity that is used to
authenticate the conveyed Evidence and establishes an attestation endpoint context. 
The attestation identity is often established as
part of a manufacturing process that embeds 
identity credentials in the entity that implements an Attester.

\item	{\em Verifier}:
The Verifier accepts Endorsements (from Endorsers) and Evidence (from the Attester) 
then conveys Attestation Results to one or more Relying Parties. 
The Verifier must evaluate the received Endorsements and Evidences
against the internal {\em appraisal policies} chosen or configured by the owner
of the Verifier~\cite{CokerGuttman2011}.
The Attestation Service Provider (ASP) 
is typically the actor which implements the Verifier role.

\item	{\em Relying Party}:
The Relying Party (RP) role is implemented by a resource manager 
that accepts Attestation Results from a Verifier. 
The Relying Party trusts the Verifier to correctly evaluate 
attestation Evidence and Policies, and to produce a correct
{\em Attestation Result}. 
Thus, we assume that the RP and the Verifier has a business relationship
(e.g. see the SAML2.0~\cite{SAMLcore} model for 
a similar business relationship assumption)
or some other basis for trusting one another. 
The Relying Party may further evaluate Attestation Results 
according to Policies it may receive from an Owner.
The Relying Party may take actions based on the evaluation of the Attestation Results.

\item	{\em  Endorser}: 
An Endorser role is typically implemented by a supply chain entity 
that creates reference Endorsements 
(i.e., claims, values or measurements that are known to be authentic). 
Endorsements contain assertions about the device's intrinsic
trustworthiness and correctness properties. 
Endorsers implement manufacturing, 
productization, or other techniques that establish
the trustworthiness properties of the Attesting Environment.
This is shown as flows~(a) and (b) in
Figure~\ref{fig:TCG-Canonical-Model}.

\item	{\em Owner of Verifier}:  
The Verifier Owner role has policy oversight for the Verifier. 
It generates Appraisal Policy for Evidence and conveys
the policy to the Verifier. 
The Verifier Owner sets policy for acceptable (or unacceptable) Evidence and Endorsements
that may be supplied by Attesters and Endorsers respectively. 

The policies determine the trustworthiness state of the Attester and
how best to represent the state to Relying Parties in the form of Attestation Results.
The Verifier Owner manages Endorsements supplied by Endorsers and may maintain a database of acceptable and/or
unacceptable Endorsements. 
The Verifier Owner authenticates Verifiers and maintains 
lists of trustworthy Endorsers, 
peer Verifiers and Relying Parties with which the Verifier might interact.

\item	{\em Owner of Relying Party}: 
The Relying Party (RP) Owner role has policy oversight for the Relying Party (RP). 
The RP-Owner sets appraisal policy regarding acceptable (or unacceptable) 
Attestation Results about an Attester that was produced by a Verifier. 
The RP-Owner sets appraisal policies on the Relying Party 
that authorizes use of Attestation Results in the context of
the relevant services, management consoles, network equipment, 
an enforcement policies used by the Relying Party. 
The Relying Party Owner authenticates the Relying Party and 
maintains lists of trustworthy Verifiers and peer 
Relying Parties with which the Relying Party might interact.

\item	{\em Evidence}:
The Attestation Evidence is a role message containing assertions from the Attester role. 
Evidence should have freshness and
recentness claims that help establish Evidence relevance. 
For example, a Verifier supplies a nonce that can be
included with the Evidence supplied by the Attester. 
Evidence typically describes the state of the device or entity.
Normally, Evidence is collected in response to a request (e.g. challenge from Verifier). 

Evidence may also describe historical device states
(e.g. the state of the Attester during initial boot). 
It may also describe operational states that are dynamic and
likely to change from one request to the next. 
Attestation protocols may be helpful in providing timing context for
correct evaluation of Evidence that is highly dynamic.

\item	{\em Endorsements}:
Endorsement structures contain reference {\em Claims}
that are signed by an entity performing the Endorser role
(e.g. supply-chain entity or manufacturer of the target device).
Endorsements are reference values that may be used by Owners to form attestation Policies.

Some endorsements may be considered ``intrinsic'' in that
they convey static trustworthiness properties relating to a given actor 
(e.g., device, environment, component, TCB, layer, RoT, or entity).
These may exist as part of the design, implementation, 
validation and manufacture of that actor implementation. 

An Endorser (e.g. manufacturer) may assert immutable and intrinsic claims 
in its Endorsements,
which then allows the Verifier to carry-out appraisal of the Attester (e.g. device)
without requiring Attester reporting
beyond simple authentication.

\end{itemize}

\subsection{Summary of an attestation event}
\label{subsec:AttestationEvent}

Figure~\ref{fig:TCG-Canonical-Model} 
illustrates the canonical attestation model. 
When an Attester (e.g.  target device) seeks to 
perform an action at the Relying Party 
(e.g. access resources or services controlled by the Relying Party) 
the Attester must first be evaluated by the Verifier. 
Among its inputs,
the Verifier obtains endorsements 
from the Endorser (e.g. device manufacturer)
in flow~(a) of Figure~\ref{fig:TCG-Canonical-Model}.
Prior to allowing any entity to be evaluated by the Verifier, 
the Owner of the Verifier must first configure 
a number of appraisal policies into the Verifier 
for evaluating Evidences. 
The policies are use-case specific but may require other 
information about the Attester (or User) to be furnished to the Verifier. 
This is shown in Step~1 of Figure~\ref{fig:TCG-Canonical-Model}.
Similarly, in Step~2 the owner of the Relying Party (e.g. resource or service)
must configure a number of Appraisal Policies for Attestation Results into the Relying Party.

When the Attester requests access to the resources at the Relying Party (Step~3),
it will be redirected to the Verifier (Step~4) -- the understanding being
that the Attester must deliver attestation Evidence to the Verifier.
Included here are the endorsement(s) that the Attester obtained previously
from the Endorser (flow(b) of Figure~\ref{fig:TCG-Canonical-Model}). 
The flow represented by Step~3 may be multi-round
and may include a nonce challenge that the Attester
must include in its computation of the Evidence as a means to establish freshness.

After verification and appraisal of the Attester completes, 
the Verifier delivers the Attestation Result to the Relying Party in Step~5.
The Relying Party in its turn must evaluate the Result
against its own policies (set previously in Step~2).
If the Relying Party is satisfied with its evaluation of 
the Attestation Result regarding the Attester,
it will provide the Attester with permission to complete the action
it seeks to perform (e.g. access resources at the RP).

\subsection{Variations in the Attestation Flows}
\label{subsec:PassportModel}

\begin{figure}[t]
\centering
\includegraphics[width=1.0\textwidth, trim={0.0cm 0.0cm 0.0cm 0.0cm}, clip]{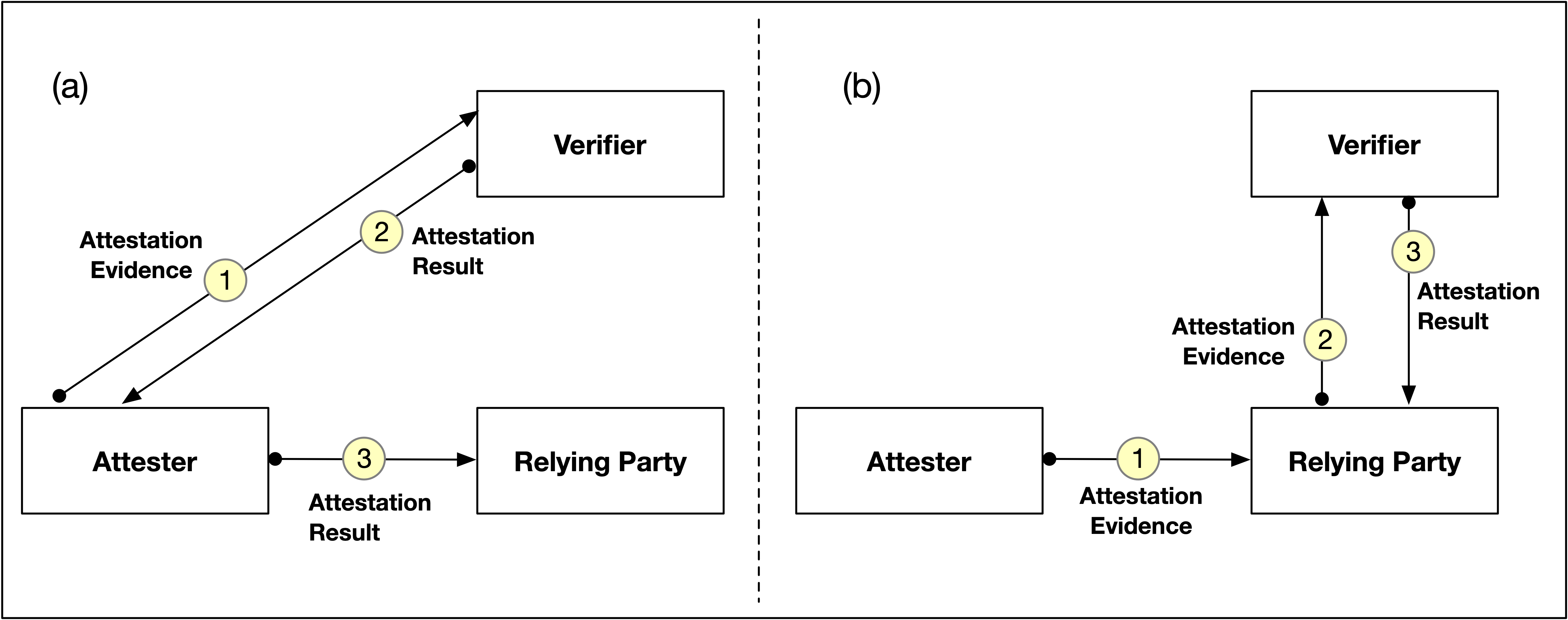}
\caption{Two variations in attestation flows: (a) The Passport flow, and (b) Background-check flow (after~\cite{rats-arch-02})}
\label{fig:FlowVariations}
\end{figure}

There are various possible variations to the 
message flows shown in Figure~\ref{fig:TCG-Canonical-Model}.
These variations may be useful and applicable to use-cases
where certain constraints are present 
(e.g. IoT device with minimal computing power, 
devices with limited connectivity, etc.).

Two (2) variations are shown in Figure~\ref{fig:FlowVariations}.
In the first case in Figure~\ref{fig:FlowVariations}(a), 
the Attester delivers its Attestation Evidence to
the Verifier as before.
However, the Verifier returns the signed Attestation Result to the Attester,
which then wields it to the Relying Party.
This variation is akin to the ``front-channel'' flow
in the Web-Browser Single Sign-On (Web-SSO)~\cite{SAMLcore} model
based on a mediated authentication service by a trusted third-party~\cite{Hardjono2019-IEEEcomms-article}.
Here the Attester's task is to convey unchanged
the (signed) Attestation Results produced by the Verifier.
This flow is referred to as the ``passport flow'' in~\cite{IETF-RATSWG}
because the Attester is wielding the Attestation Results
in the manner of a signed passport of permit.

In the second variation shown in Figure~\ref{fig:FlowVariations}(b), 
the Attester delivers its attestation Evidence direct
to the Relying Party (i.e. resource manager).
Being a reliant party -- reliant on the Verifier to evaluate attestation Evidences --
the Relying Party simply forwards the Evidence to the Verifier.
After the Verifier completes appraisal of the attestation Evidence,
it returns the Attestation Results to the Relying Party directly.
This flow is referred to as the ``background check flow'' in~\cite{IETF-RATSWG}
-- or ``back-channel'' in SAML2.0 literature~\cite{SAMLcore} --
because the delivery of the Attestation Result occurs between the Verifier
and the Relying Party over a one-to-one channel without the assistance of the Attester.

\begin{figure}[t]
\centering
\includegraphics[width=1.0\textwidth, trim={0.0cm 0.0cm 0.0cm 0.0cm}, clip]{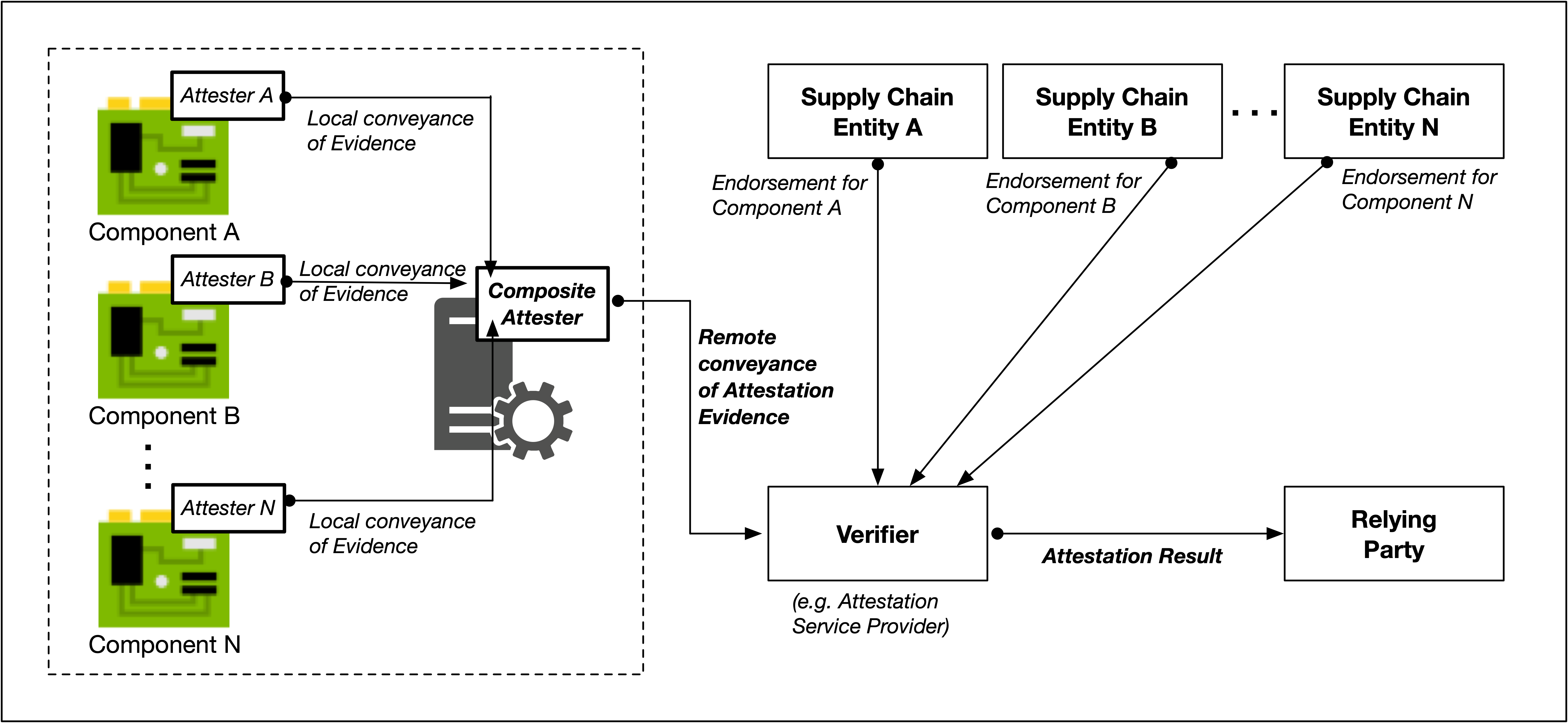}
\caption{Example of Composite Device Attestations}
\label{fig:CompositeAttestations}
\end{figure}

\subsection{Composite Attestations}
\label{subsec:CompositeAttestations}

In some cases, an attestation Evidence yielded by an Attester may in fact consists of
other Evidence (e.g. from other local components) collated by 
that Attester (Figure~\ref{fig:CompositeAttestations}).

We refer to this kind of attester as the {\em Composite Device Attester}
and the evidence as {\em Composite Device Evidence}.
In a composite attester scenario, we assume local components have attestation capabilities 
that generate evidence. 
This evidence is conveyed locally to a lead attester that assembles 
the various sets of evidences, 
possibly including evidence that it directly collects as well. 
Subsequently,
the composite device lead attester conveys composite evidence to the relevant Verifier.
The composite attester may assert a claim that it was the entity that assembled a piece of component evidence and 
include this assertion in the composite device evidence it supplies.

\subsection{Layered Attestations}
\label{sec:LayeredAttestations}

\begin{figure}[t]
\centering
\includegraphics[width=1.0\textwidth, trim={0.0cm 0.0cm 0.0cm 0.0cm}, clip]{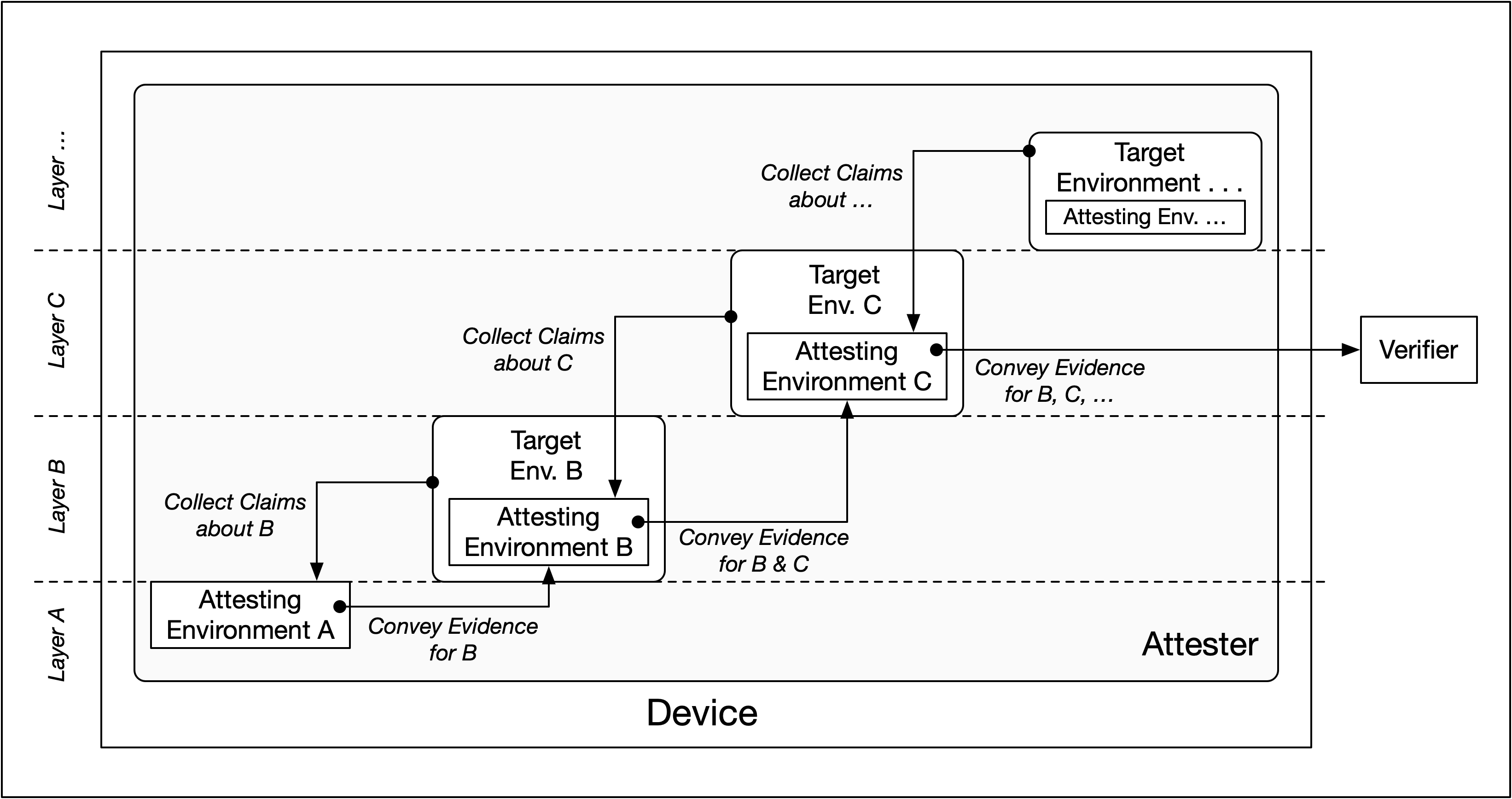}
\caption{Overview of the concept of the Layered Device Attestations}
\label{fig:LayeredAttest}
\end{figure}

Another mode of deployment for the attestation architecture
is in the appraisal of software (firmware) modules relating
to the boot-up sequence within a given device.
As mentioned previously,
the Trusted Computing Group (TCG) defined a number of hardware-based ``roots-of-trust'' (RoT)
related to the TPM chip~\cite{TPM2003Design}.
The idea was that because the TPM hardware was tamper-resistant
and provided shielded memory and storage,
these features could be used as a root-of-trust
for ensuring that a TPM-enabled device could boot-up safely and correctly.

However, the TPM-based approach may not be suitable for various other use-cases,
and not all devices can suitably support a TPM.
Constrained devices such as IoT devices (Internet of Things) -- such a low-power sensors and tiny low-cost devices --
may not be able to support a TPM or any dedicate security processor.
More importantly,
not every device architecture maps onto the assumptions underlying the TPM design.
In some cases, once the device completes loading
the low-level software -- whose integrity can be protected by keeping a hash 
of the code in the registers of the TPM --
there comes a point in the boot-sequence where the TPM ceases to be (i.e. unable to be) 
the root-of-trust for the next piece of software being loaded.
That is, there needs to be a variant of the attestation model which
can be aided in its initial phases by the use of some 
hardware-based functions and relevant manufacturer endorsements,
but which would require it to be reliant on attestations
by other pieces of software in the boot-sequence.
We refer to this generically as {\em layered attestations} (Figure~\ref{fig:LayeredAttest}).

One specific example of layered attestations can be found
in the {\em Robust Internet-of-Things} (RIoT) architecture~\cite{England2016RIOT}.
The approach employs a combination of a device-secret
that is set by the device manufacturer during production (e.g. fusing during manufacturing).
The core idea is to use the device secret and a keyed hash function
to derive other secrets (e.g. keys) to be used by the next layer in the boot-sequence.

Although a detailed discussion of layered attestations
is beyond the scope of the current work,
there are a number of desirable properties
for layered attestations (see Section~5 of~\cite{HardjonoSmith2019c} for an extensive discusion).
First, each layer in the sequence must be 
unambiguously distinguishable (e.g. using a key derivation scheme
that provides a unique key for each layer).
Secondly, the next layer must be ``inspectable''
be the current layer.
Inspection may simply be to compute a hash value
for computation of a layer identity or may involve more
rigorous proofs of integrity.
Thirdly,
there must be a way to achieve layer sequencing -- which
may different for each type of device.
Finally,
there must be a way for a layer to provide Attestation Evidence
of itself that includes evidences for all previous layers
in the sequence.
Trust in a current layer depends on the trustworthiness of all previous layers. 
Consequently, a Verifier of layered attestation must evaluate Attestation Evidence 
of all the dependent layers before it can reason about trust in the current layer.
The set of layered evidence must therefore be communicated
within the evidence flow emanating from the Attester.

The TCG has formalized and standardized the notion of layered attestations
in the {\em Device Identity Composition Engine} (DICE) 
specifications~\cite{TCG-DICE-Implicit-2018,TCG-DICE-Symmetric-2020}.
This standardization process is important not only
from a device-manufacturers perspective (and other supply-chain entities),
but also from the perspective of the various
service providers (e.g. ASPs and Relying Parties)
that together with the supply-chain entities
form the ecosystem that supports interoperable implementations
of the attestation architecture. 
Currently, the DICE approach has been implemented
for hardware intended for cloud platforms
(e.g. Project Cerberus~\cite{Kelly-Cerberus-2017}).


\section{Attestation of Nodes in Blockchain Networks}
\label{sec:AttestationsNodes}

In this section we explore the application of the canonical attestation architecture in Section~\ref{sec:TCGAttestation}
to the broad case of blockchain networks. 
We pay close attention to {\em permissioned} (private) blockchain networks
as a means to constrain the scope the problem.
We envisage that attestation architectures and technologies discussed in the current work
may be of interest in the first instance 
to communities arranged as consortiums seeking to employ blockchain technology
and distributed ledger technologies (DLT) generally to solve a specific problem in the community.
Currently, several organizations have embarked on creating the technology and infrastructures
to effect permissioned blockchains 
(e.g. R3/Corda~\cite{R3-website}, 
TradeLens~\cite{Miller2018},
PharmaLedger~\cite{Morris2020}).
In the financial industry,
several organizations are exploring the notion of private/permissioned
blockchains for the purpose of increasing the efficiency of business 
settlements within their network~\cite{Castillo2020}.

\begin{figure}[t]
\centering
\includegraphics[width=0.8\textwidth, trim={0.0cm 0.0cm 0.0cm 0.0cm}, clip]{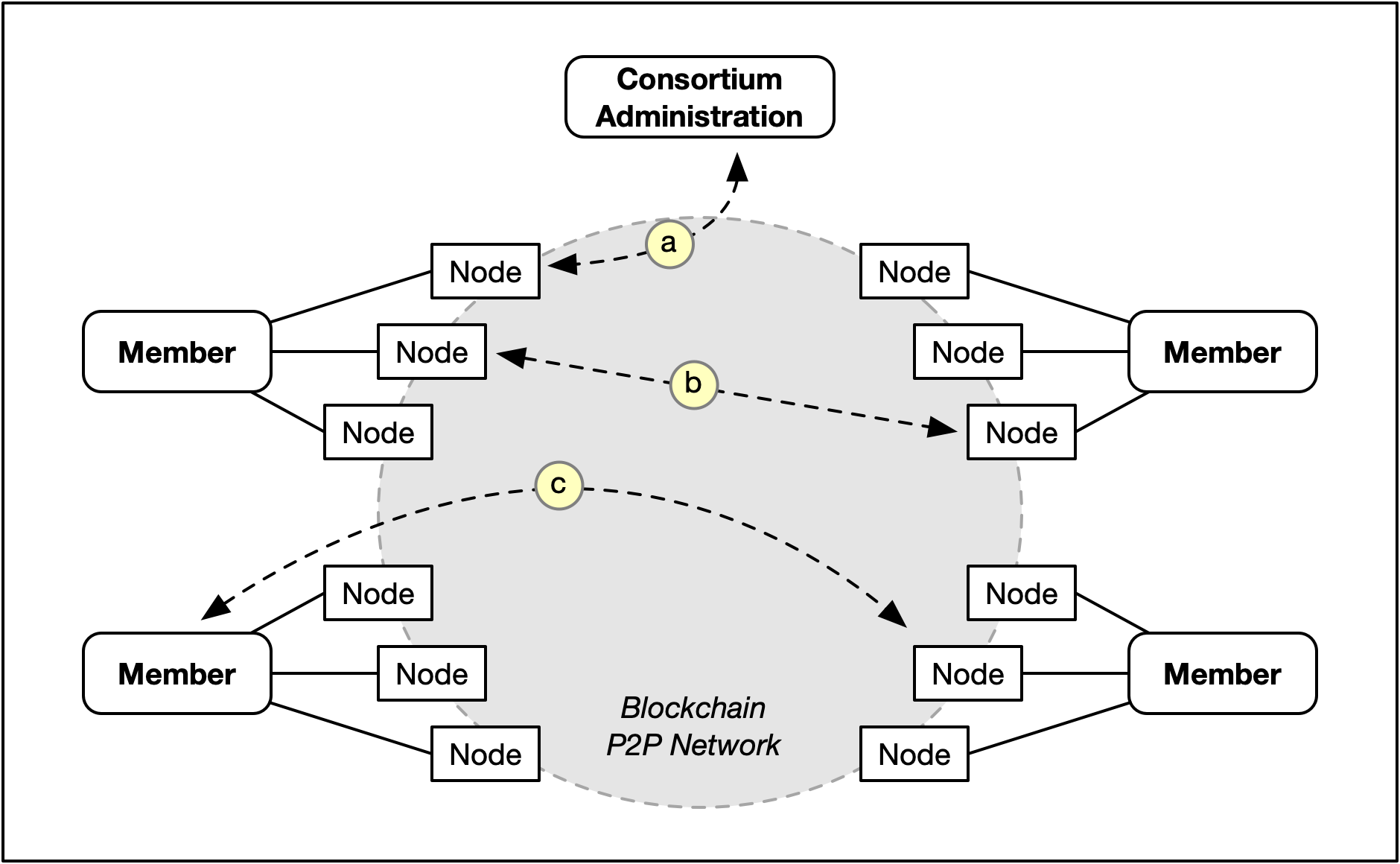}
\caption{Overview of a consortium arrangement of a permissioned blockchain network}
\label{fig:consortium}
\end{figure}

There are several high-level desirable features for
nodes in a blockchain network:
\begin{itemize}

\item	{\em Independence of nodes in verifying other nodes}:
A node must be able to take-on the role of 
a Verifier of the attestation Evidence generated by peer nodes. 
This means that nodes must have awareness of the identity of its peer nodes.
Similarly,
a node must be able to generate a signed Attestation Evidence
as required (e.g. when demanded by its owner,
by the consortium, by peer nodes,
or as required by the consensus-protocol in the network).

\item	{\em Persistence of transaction signing key over reboots}:
Since a node's transaction signing key may be used
in some consensus-protocols to receive remunerations (e.g. BTCs, ``gas''),
this private-public key pair must be persistent (i.e. not lost)
across reboots of the node-device.
This must be true independent of the hardware/software
implementation of the node.

\item	{\em Inaccessible transaction signing keys when in unapproved configuration}:
Since a node must only operate in a configuration and state that
is known to and approved by the node's owner,
the transaction signing key(s) must not be usable or accessible
by the node if it is in an ``unhealthy'' (i.e. unapproved) state.
Among others,
this is to prevent malware (i.e. viruses) from using
transaction signing keys to perform unauthorized transactions,
thereby harming its owner.

\item	{\em Observance and enforcement of governance polices}:
In a consortium arrangement,
there must be a way for the consortium organization
to mandate (force) observance by a node
of the consortium-wide policies and operating rules.
There are various mechanisms that can achieve this effect,
one of which is to use Attestation Results
as one of the inputs into the consensus-protocol
(e.g. if a node is not in one of several configurations
approved by the consortium,
then the node will never be selected 
to forge new blocks in a Proof of Stake protocol).

\end{itemize}

Figure~\ref{fig:consortium} illustrates a number of attestation flows
that may occur in a consortium-based permissioned blockchain network.
The flow~(a) in Figure~\ref{fig:consortium} illustrates situations
where the consortium governance administration seeks
to verify the attestations produced by nodes belonging to members.
This ``right to verify'' may be enshrined within the governance operating contracts
of the consortium.
In flow~(b) nodes are performing mutual attestations of peer nodes
in an independent manner (i.e. independent of any centralized entity
such as the consortium administration entity).
In flow~(c),
a member may employ its own verifier to evaluate nodes belonging
to other members in the consortium.

\begin{figure}[t]
\centering
\includegraphics[width=1.0\textwidth, trim={0.0cm 0.0cm 0.0cm 0.0cm}, clip]{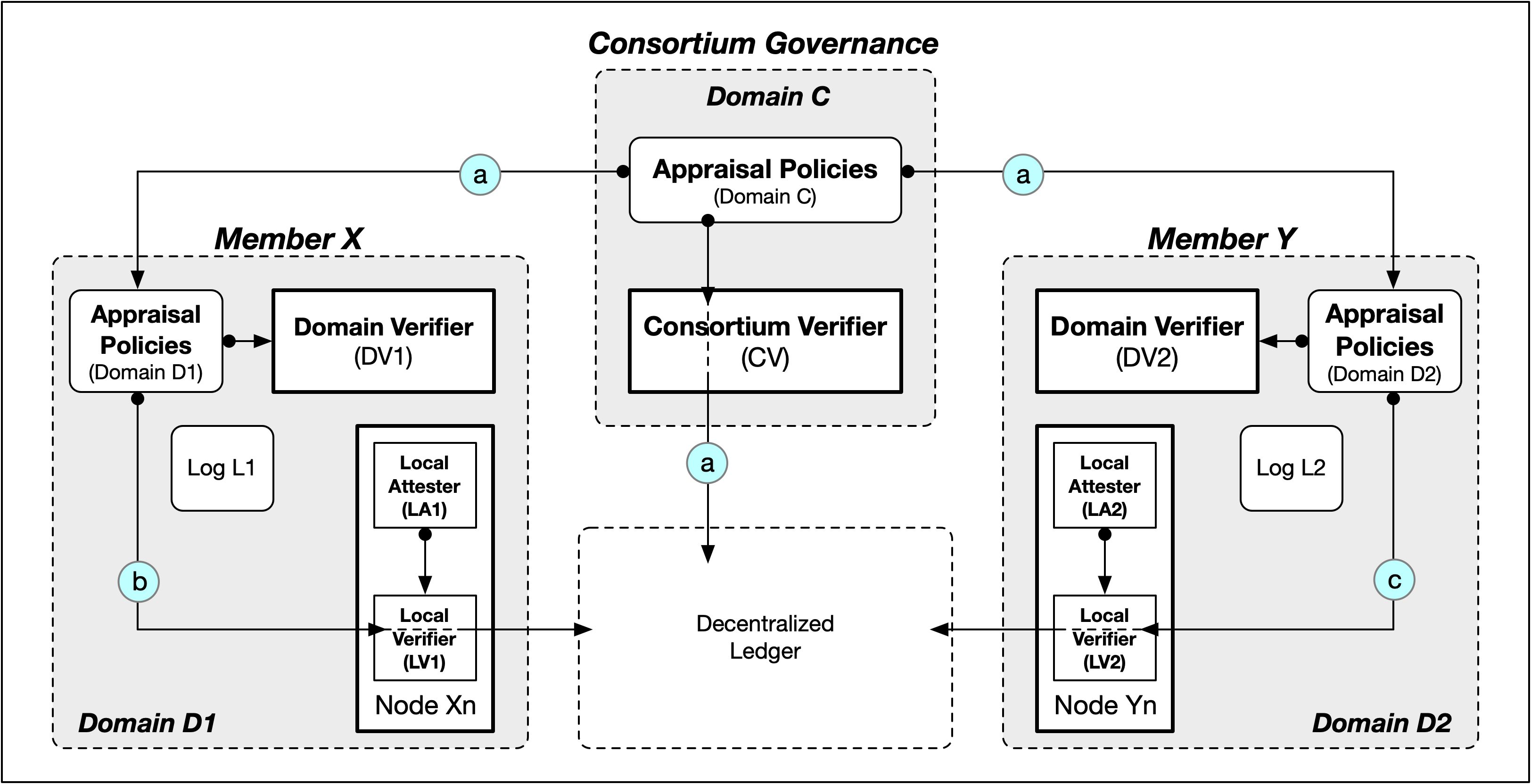}
\caption{Architecture for attestations in a permissioned blockchain network}
\label{fig:BasicsNodeAttestations}
\end{figure}

\subsection{Domains, Nodes and Functions}
\label{subsec:}

Following from the general entities and roles defined in Section~\ref{subsec:TCG-Attestation-entities},
in the following we apply these to the entities that
participate in a blockchain network 
(Figure~\ref{fig:BasicsNodeAttestations}):
\begin{itemize}

\item	{\em Consortium Verifier}: The consortium as a community owns and operates
one or more of its own Verifiers (e.g. Attestation Service Provider) 
for the purpose of appraising Evidence conveyed by nodes in the network
against the consortium's Appraisal Policies.

\item	{\em Consortium-wide Appraisal Policies}: Part of the governance of the community
is the establishment of a shared set of appraisal policies 
for Attestation Evidences and Attestation Results.
Step~(a) of Figure~\ref{fig:BasicsNodeAttestations} illustrates
the conveyance of the consortium-wide appraisal policies
to the domains (e.g. the policy store in the management console of domain owner).
The intent of Step~(a) is to denote that the consortium-wide appraisal policies
must be communicated down to the Local Verifier in each node in that domain.
Copies of the Appraisal Policies are also recorded on the ledger
(i.e. digest or hash of the policies)
to allow for future independent audit,
such as comparing member policies against the
member-agreed baseline policies of the consortium.

\item	{\em Member Verifier}: Each member owns and operates
one or more or its own Domain Verifier.
Thus, for example,
in Figure~\ref{fig:BasicsNodeAttestations} a Member~X 
owns the Domain Verifier DV1 in Domain~D1.
Each member also operates a local protected {\em Log},
which is only accessible by nodes and other entities in the domain.

\item	{\em Member Appraisal Policies}: 
Each member may set its own appraisal policies for 
Attestation Evidences and Attestation Results
for its domain.
Ideally, member-level policies should not conflict with
consortium-level policies.
The consortium has the opportunity
to use independent audit and compliance entity 
to ensure that correctness evaluation can be performed.
Step~(b) and Step~(c) in Figure~\ref{fig:BasicsNodeAttestations} illustrates
the conveyance of the consortium-wide appraisal policies
and the domain specific appraisal policies
to the nodes of the corresponding member
(with digests recorded on the ledger).

\item	{\em Node Local Verifier and Local Attester}: 
Each node in the blockchain network 
implements a Local Verifier (LV) and Local Attester (LA).
The Local Attester creates attestation Evidence about the node
and conveys the evidence to a Verifier
(i.e. its own Local Verifier,
its Domain Verifier,
the Consortium Verifier,
or another node's Local Verifier).

\item	{\em Audit Log}: A domain maintains an audit log system
as means to store and retain attestation Evidences
(from devices within its domain)
and Attestation Results coming from the various Node Local Verifiers
as well as the Consortium Verifier (CV).
The entries of the local log
(i.e. hash of entries) can also be recorded on the ledger.
This permits independent audit and compliance entities to 
check the results are reasonable appraisal results.

\end{itemize}

There are two broad categories of scenarios that can benefit from attestations.
In the first scenario, a Local Attester in a given node
conveys attestation Evidence to a Verifier that is
either located in its home domain or in the consortium domain.
In the second scenario,
a Local Attester in a given node
conveys attestation Evidence to a Verifier located in another node
in a peer-to-peer fashion.
We discuss these scenarios below.

\begin{figure}[t]
\centering
\includegraphics[width=0.8\textwidth, trim={0.0cm 0.0cm 0.0cm 0.0cm}, clip]{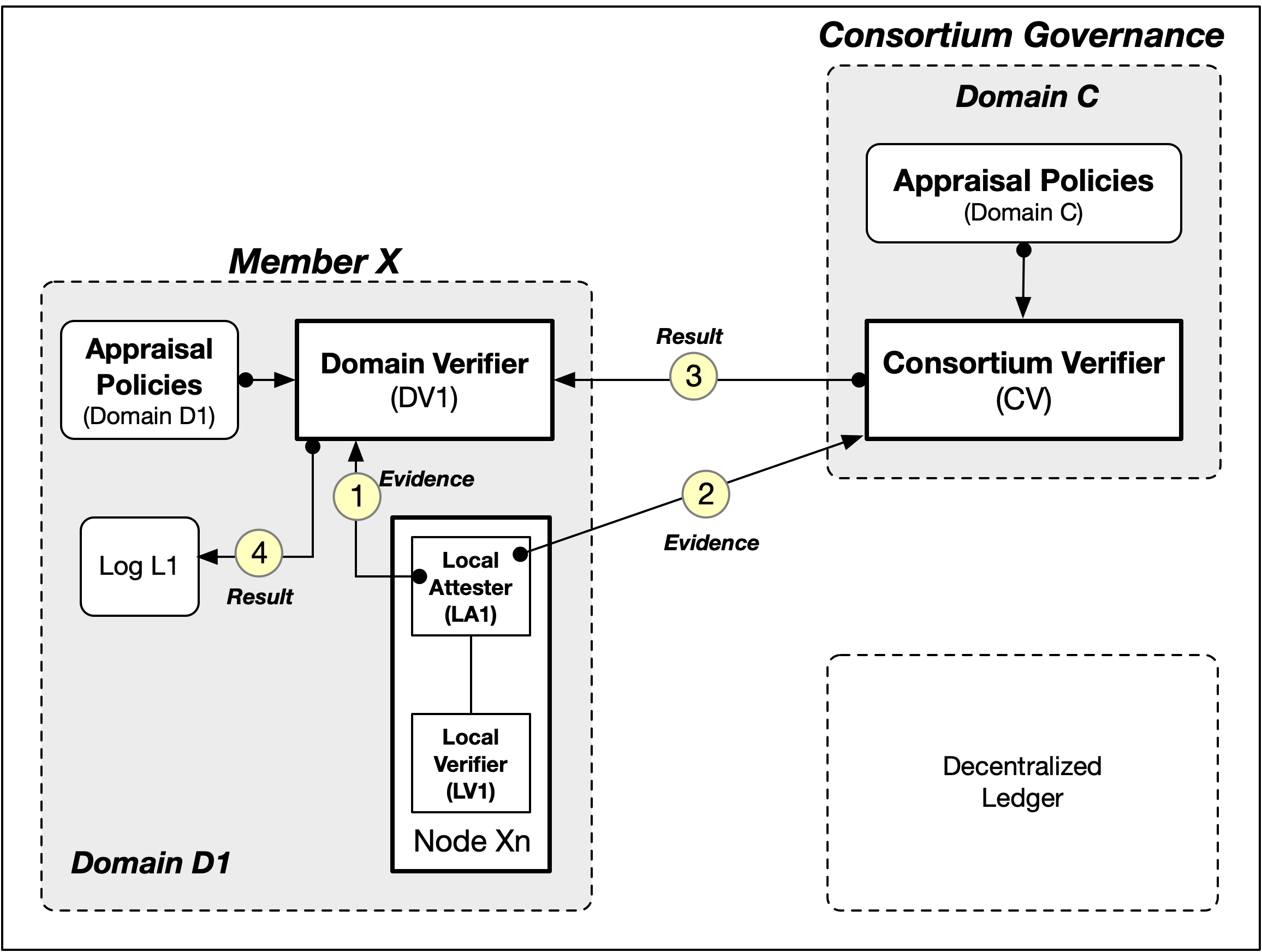}
\caption{Attestations of a node by its Domain Verifier}
\label{fig:LocalAttestations}
\end{figure}

\subsection{Appraisal by Domain Verifiers}
\label{subsec:localdomainattest}

There are a number of use-cases related to manageability of nodes/devices
that may benefit from the ability for a node to
convey Evidence regarding its current configuration
and other computations state.
We use Figure~\ref{fig:LocalAttestations} to illustrate.

In Step~(1) and Step~(2) of Figure~\ref{fig:LocalAttestations}
the Local Attester (LA1) in the node
conveys attestation evidence to (i) its own
Domain Verifier (DV1) and (ii) to the Consortium Verifier (CV).
Depending on the specific-use case,
these two evidences may differ.
Thus, for example,
the Domain Verifier may be concerned about both the health of its
node and other system attributes of its node
(e.g. did a GPU card just malfunctioned).
The Evidence conveyed by the Local Attester (LA1)
to the Consortium Verifier (CV)
may include information that is of interest
to the consortium administration.
For example,
the Consortium Verifier may be seeking status information
regarding the versions of the firmware and software
on the node,
when the last patch was installed, and so on.
The goal of the consortium's appraisal policies
is to ensure that members and their nodes comply
to the operating rules of the consortium organization
as defined in its legal trust framework and other related membership contracts.

Note that these two areas of interest -- of the Domain Verifier on one hand
and the Consortium Verifier on the other hand --
are complementary to each other.
The understanding here is that the survival of the permission
blockchain network as a whole
is a function of the survival of individual nodes in that network. 
When nodes are ``healthy'' then the network is also healthy.

It is worthwhile to note that the Relying Parties (RP)
in these two scenarios are the Domain Owner (i.e. node owner)
and the Consortium Administration respectively.
Thus, in Step~(3) and Step~(4) the attestation results
are conveyed by the Consortium Verifier and by the Domain Verifier (DV1) respectively.
The Domain Owner as a relying party
to both of these attestation results may take action based on the received results.
For example,
it could update the firmware of the node,
to bring the node offline, and so on.
Thus, the owner's foremost concern could be the visibility into the state of its nodes
and maintaining the security and integrity of these nodes.

\subsection{Appraisal by Peer Nodes}
\label{subsec:peered-mutual}

One of the key tenets of decentralized computing as exemplified by blockchain systems
such as Bitcoin is the autonomy of nodes
in performing some computations 
(e.g. Proof-of-Work~\cite{Bitcoin}, Proof-of-Stake~\cite{Ethereum-POS-2019}, etc.).
To this end,
ideally nodes must be able to appraise other nodes
as part of a consensus-making protocol.
The ability to provide attestation evidence (e.g. regarding a node's current configuration)
enhances the acceptability of the outcomes of peered computations
such as PoW and PoS.
Other participants in the network obtain some degree of assurance that
fairness has been maintained (e.g. that nodes have equal hash-powers or hash-rates).
This is especially important for consortium organizations whose members may be competitors.

Another important use-case
pertains to nodes that act as gateways between two distinct blockchain networks~\cite{HardjonoSmith2019c}.
Here, the goal of gateways is the establishment of trust for scenarios involving high value
transactions.
For certain types of virtual assets (e.g. proof of legal ownerships of real assets)
a change of legal ownership effected on a blockchain may necessitate that 
the evidence be moved from one blockchain to another
(e.g. from the seller's blockchain to the buyer's preferred blockchain).
This movability of virtual assets across blockchain systems
is crucial for the scalability of blockchains as an economic medium for business transactions at a global scale.

\begin{figure}[t]
\centering
\includegraphics[width=1.0\textwidth, trim={0.0cm 0.0cm 0.0cm 0.0cm}, clip]{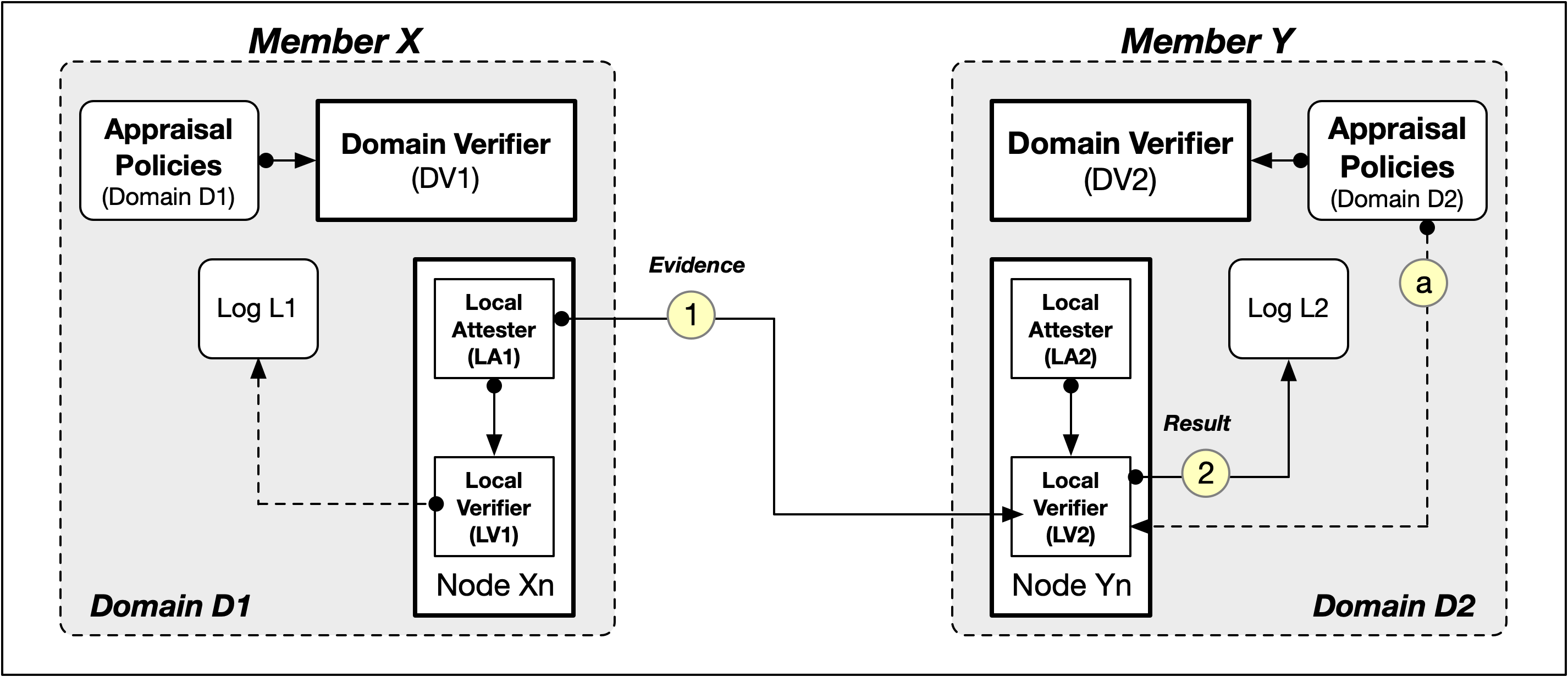}
\caption{Appraisal of attestations by peer nodes}
\label{fig:PeeredAttestations}
\end{figure}

The appraisal by peer nodes is represented by Figure~\ref{fig:PeeredAttestations}.
Here a node $X_n$ is required to provide attestation Evidence to other
nodes in the network (e.g. node $Y_n$) as part of consensus-making
(e.g. node $X_n$ to be selected to forge new blocks in PoS~\cite{Ethereum-POS-2019}).
In Step~(1) of Figure~\ref{fig:PeeredAttestations}
the attestation evidence from node $X_n$ is conveyed
to the Local Verifier (LV2) within node $Y_n$.
Now, LV2 will evaluate the received evidence
based on the appraisal policies (for Evidences and Attestation Results)
which it possesses. These policies were previously configured
by the Domain Verifier owner (node owner) in Step~(a) of Figure~\ref{fig:PeeredAttestations}.
If the results of the attestations conforms to the appraisal policies,
the node $Y_n$ (as its own Relying Party) may then
take action (e.g. confirm proposed new blocks in PoW).

\section{Attestation of Virtualized Nodes: Future Challenges}
\label{sec:AttestVirtualizedNodes}

Since the emergence of the Bitcoin system~\cite{Bitcoin} in 2008,
there has been significant development and departure from the original
Bitcoin conception of the topology of nodes (i.e. mining nodes).
One key idea in Bitcoin is that the use of physically-separate nodes (i.e. mining rigs) -- each with
its own copy of the full ledger --
would provide a degree of resilience of the network to attackers
who wish to skew or compromise the network.
An attacker would need to successfully attack a majority (e.g. 51 percent)
of the physically-separate nodes in order to compromise the network as a whole~\cite{EyalSirer2014,GervaisKarame2014}.
The Ethereum system~\cite{Buterin2014} represented a break from classic Bitcoin topology
by expanding the programmability of the blockchain through the use
of ``smart contracts'' that operate within the Ethereum Virtual Machine (EVM).
The EVM is essentially stack machine~\cite{StackMachine-Wiki} that operates in a virtual space
made available on the nodes of Ethereum.

\subsection{Cloud Computing, CaaS and BaaS}
\label{sec:CloudComputing}

Given the complexity of operating nodes
and the resources needed to maintain a network of nodes,
it is reasonable to expect that the nascent  
{\em virtual assets}~\cite{FATF-Recommendation15-2018,HardjonoLipton2020a} 
industry will look to cloud computing as a means
to increase scale while reducing operational costs.
Indeed, currently new forms of cloud-based offerings have begun to emerge
marketed to various blockchain-based use cases.
These offerings range from
{\em Container as a Service} (CaaS) to {\em Blockchain as a Service} (BaaS).
In the CaaS case, the node-owner can
create an image (e.g. Docker image) of the node and have it execute 
in the CaaS infrastructure (e.g. IBM cloud~\cite{IBM-Blockchain-2019}).
In the BaaS case,
the entire blockchain network can be hosted on a third-party infrastructure
(e.g. Microsoft Azure BaaS~\cite{Lardinois2019}).

However,
several challenges remain to be addressed with regards to 
the CaaS and BaaS models for blockchain networks.
Some of the challenges related to nodes implemented in a BaaS platform include:
(i) the security of cryptographic keys of nodes in the cloud;
(ii) the integrity -- and in some cases the confidentiality -- of the ledger data held by nodes;
(iii) the secure migration of nodes -- or processes implementing the node -- from one
virtualized stack to another;
(iv) malicious interference by adjacent processes in multi-tenant deployments;
(v) geographic diversity of the nodes;
and so on.

\subsection{Geographic diversity of nodes}
\label{sec:Geographicdiversity}

Historically, geographic diversity has not been reliably enforced as part of a blockchain network. 
For example, the bitcoin Proof-of-Oork (PoW) focuses
mainly on how how quickly the miner can solve the PoW cryptographic puzzle. 
For the resilience of the network,
ideally nodes (miners) should be geographically spread,
and geo-politically diverse.
Geopolitical diversity has ramifications on 
the stability of the value of the virtual assets 
transacted on the blockchain (e.g. see~\cite{TradecoinSciAm2018}).
The attestation architecture we have described above
permits nodes/devices to provide evidence of its geolocation.
For example,
the work of~\cite{IETF-draft-ietf-rats-eat-03} includes the ability to
report location coordinates (latitude, longitude and altitude)
of the attester device.
In turn,
this can be reinforced with geo-fence policies relevant to the specific deployment scenario.

We believe that both geo-location evidence and geo-fence policy compliance
should in fact be integrated into the consensus protocol of the blockchain network
(e.g. PoW, PoS or future variants)
where the consensus protocol enforces geo-diversity.
This should be true including for nodes implemented using CaaS or BaaS technology.
Using the PoS example,
a node's eligibility factors -- to be selected as the validator node to forge the next block --
should include the geo-location of the node and the geo-locations of the population
of the other eligible nodes.
Using the peer-appraisal approach outlined in Section~\ref{subsec:peered-mutual},
a community of nodes could collectively self-enforce this geo-diversity requirement
as part of their consensus-making algorithm.

A similar notion in the context of IP routing -- referred to as {\em trusted path routing} --
has been proposed in~\cite{IETF-draft-voit-rats-trusted-path-routing-01},
where routers (i.e. routing nodes)
in a traditional IP network employ attestation of peers to 
exclude routers whose attestation evidence does not
meet a policy agreed upon by the community of nodes. 
Although trusted path routing envisages a distributed attestation appraisal solution 
that approaches a distributed consensus algorithm,
more work could be directed at developing
an ``equivalent'' of consensus algorithm but for routers eligible to be in the trusted-path
(i.e. route selection versus mining/forging blocks).

\subsection{Integrity of Cloud Platforms}
\label{sec:CloudIntegrity}

We believe that a secure attestations capability for nodes 
implemented on a BaaS (or CaaS) platform
represents an important factor for the business value-proposition of the BaaS model.
Attestations capability are needed at different layers of the virtualization stack
according to the corresponding ``consumer'' (Verifier and Relying Party) 
of the attestation information.
Thus, for example,
the BaaS provider needs visibility into the state of the platform as a whole,
while the node-owner needs visibility into the integrity of its node.
Counter-parties in transactions may wish to see attestation results
for nodes running on the BaaS platform, and so on.

For cloud providers generally, there are a number of challenges in providing
a reliable and safe virtual computing infrastructures~\cite{NIST-800-193}.
There needs to be a way to validate the integrity of all firmware updates
as a follow-up after the {\em first-instruction integrity} has been completed~\cite{CSA2018}.
Thus, after the integrity of the first code or data loaded (e.g. from mutable non-volatile media)
has been verified (e.g. by either the cloud provider or the manufacturer),
the integrity of all firmware updates must also be achieved in a verifiable manner~\cite{England2016RIOT}.
Achieving this higher level of integrity implies that there could be several 
roots of trust (RoT) or chains of trust (CoT) that are integral to the platform.

Furthermore, there needs to be a way to {\em detect unauthorized access or corruption}
to the software or firmwares during operations,
and then take recourse to remediate this problem -- possibly
in an automated or semi-automated manner.
There needs to be a way to {\em restore firmware to state of integrity} in cases where
corruption has been detected~\cite{NIST-800-193}.
Given the nature of cloud data centers,
recovery must be achievable in an automated fashion without 
immediate attendance of the IT administrator.
Manual recovery must of course be supported.
Recovery in this case generally means 
automatic self-recovery of the critical-to-boot portion(s) of the firmware.

\subsection{Decentralized Roots of Trust}
\label{sec:DistributedRoots}

Another challenge pertains to the notion of 
{\em decentralized roots of trust} (d-RoT) put forward in~\cite{HardjonoSmith2019c},
where the logic of trust for a given multi-node (multi-component)
environment should be based on a decentralized 
cooperation of those nodes (components).
The patterns of roles, interaction and relationships we saw previously in Section~\ref{sec:TCGAttestation}
-- namely endorsements, attestation-evidence, appraisal and attestation-results --
must hold true within individual computer systems
(which may consists of a complex interaction of parts and components)
as well as within a group of distributed computer systems (e.g. nodes on a blockchain network).

For example, within a single computer system
a root-of-trust in a hardware CPU or a core package containing multiple cores 
may have a design that justifies it as a root-of-trust.
However, 
when put together with other components on a system bus
there maybe other roots-of-trust present
(e.g. system-on-chip (SoC), IP blocks and peripherals, etc)
that may have equal access to the system bus and therefore
can assert themselves as a root-of-trust.
In this case 
there must be a distributed mechanism for bootstrapping 
trust in the system as a whole based 
on a collective action or consensus of these roots-of-trust.
Thus, there should be (must be) be a distributed trust logic being applied 
at this {\em microcosm} level of components (``nodes'')
within one environment (the computer system as a whole).

Equally, the distributed trust logic must also apply at 
the {\em macrocosm} level consisting of multiple computer systems,
each of which are acting as a distinct node participating with other nodes within a blockchain network.
The blockchain network becomes a natural extension
of the principles and properties 
described in~\cite{HardjonoSmith2019c} (e.g. group reporting, group computation participation, etc.),
with the appropriate
attention given to constructs at the macro level,
such as correct integration with the consensus algorithms and architecture 
for establishing a d-RoT at the macro (peer-nodes) level.

\section{Areas for Innovation}
\label{sec:Innovation}

Despite the notion of device attestations nearing two-decades 
in age~\cite{TPM2003Design},
the concepts around attestations -- such as endorsements, validations and freshness --
are just recently coming into wider attention in the broader industry.
We believe more research needs to be applied,
and several areas of innovation still await the industry as a whole.
In the context of the application of the attestation architecture
to blockchain networks the following represents 
a brief list of possible areas for future innovation:
\begin{itemize}

\item	{\em  Dynamic governance-policy setting based on attestations}:
For permissioned blockchains in a consortium arrangement,
the governance-level appraisal policies should be dynamic
in that their governance parameters 
should be subject to orchestrated change 
that adapts to the shifting environment of software evolution 
and hardware replacement cycles.
This should be done in accordance with a combination of
the parameter for category type (e.g. node diversity) and 
the overall state of the population of nodes in the network.

Thus, given a node that addresses diversity (software and hardware diversity), 
if for some reason the population of nodes become increasingly non-diverse (homogeneous)
up to defined threshold, then the governance policies should change to account 
for the loss of expected diversity -- and subsequently influence 
(i.e. modify) the consensus algorithm toward stasis of an intended diversity metric. 
This could be a direct change in the parameter of the consensus algorithm 
(e.g. majority parameter raised to 70 percent of population from 51 percent),
or it could be an indirect change through new policies being pushed into the 
domain-level (member level) appraisal policies 
(e.g. Local Verifier in a node belonging to a subset of diverse nodes 
should prioritize members of that same subset
when accepting proposed consensus outcomes).
The goal should be, among others,
to incentivize members of the consortium to deploy
nodes that satisfy the diversity category (software and hardware diversity)
for the sake of the resilience of the community as a whole.

\item	{\em  Attestation for migration of containerized-nodes}:
In the context of trusted hardware used in virtualized platforms, 
one difficult challenge facing cloud data centers 
is {\em trusted migration} of containers and functions~\cite{FAAS-Wiki,Fowler-Serverless}. 
This is especially important for containerized nodes that carry 
sensitive information such as 
application-layer cryptographic keys and 
related keying material, which may need higher availability.

By design, a container is unaware of ``where'' it executes 
(i.e. the specific hardware environment -- model/version). 
However, as we have seen above, execution environment endorsements necessarily 
describe a piece hardware (with its firmware and software). 
This is crucial for reasoning about trust among multiple interrelated nodes. 
Given that attestation-derived trust reaches past the containers' presumed isolation domain, 
the problem becomes complicated by the need for 
prescribing destination hardware (to which a container is to be migrated) 
that provides ``equivalent trust'', 
but where the migration target might not exactly mimic the currently executing container. 
Properly secured migration policy expects equivalent or 
better trust in the target migration environment.

A related challenge facing trustworthy container migrations 
is how to automate the verification (comparison) between 
the current execution environment versus the target migration environment.

\item	{\em  Attestation for nodes of permissionless blockchain networks}:
To avoid confusion when discussing permissionless blockchains,
we employ the NIST definition of permissionless blockchains~\cite{NIST-80202-2018},
namely a system where all users' permissions 
are equal and not set by any administrator or consortium.
Permissionless blockchain networks are decentralized 
ledger platforms open to anyone publishing blocks, 
without needing permission from any authority (see Section 2.1 of~\cite{NIST-80202-2018}).

One possible application of attestations in permissionless blockchains
with anonymous nodes (e.g. miners)
is for the gradual establishment of a subset 
of attestation-capable nodes that periodically
report attestation-evidence at a regular basis
to the public ledger.
For simplicity, we refer to these as ``honest-nodes''.
Honest users wishing to choose to have their transactions
be processed by one or more of this subset of honest-nodes
can pre-register (i.e. self-declare) their public-keys
to these honest-nodes.
In turn,
when an honest-node searches through the list
of unprocessed transactions (e.g. in the UTXO model~\cite{UTXO-model}),
the node can choose only those new transactions 
which originated from pre-registered user public-keys,
according to the attestation policy common among the honest-nodes. 
Users can directly remunerate the honest-nodes
(who successfully created a block containing transaction
from honest-users)
by sending coins to the address of these known honest-nodes.

In effect, this creates a {\em segregation of honest-nodes}
from the broader population of anonymous nodes in the 
permissionless blockchain,
something that might be considered  a form of {\em semi-permissioned} blockchain. 
This may provide a path forward for blockchains
that today suffer from the imbalance of hash-power,
where some nodes (i.e. mining pools) control too much hash-power
and therefore introducing potential instability into the blockchain.
Such an approach is outlined in~\cite{HardjonoPentland2016}
based on the use of certain types of anonymity-preserving keys in the TPM hardware
(e.g. DAA~\cite{Brickell2004}, EPID~\cite{BrickellLi2012}).

\item	{\em  Blockchain-based Retention of Supply-Chain Endorsements}:
Beyond the challenges to ensure that endorsements are correct 
and source-authentic, there is the industry challenge 
of ensuring the continual availability and update 
of endorsements, and the availability infrastructures
and services (e.g. certificate services)
which support attestation verifications.
From a business perspective there are at least two challenges 
faced by commercial manufacturers: 
(i) a manufacturer ceases to exist, and/or
(ii) an issuing CA ceases to exist. 
A manufacturer may go out of business entirely, 
or undergo mergers with other business entities. 
At the same time, the CA that issued 
a ceased-manufacturer itself may also go out of business.
Blockchain technology -- consisting of multiple 
nodes retaining full copies of the shared ledger -- may provide 
a way to alleviate the lack of endorsement availability and 
shifting source-authenticity challenges.

The basic notion is that when a manufacturer today creates a product
and the endorsements for that product,
these endorsements could be ``registered''
on a special blockchain (referred to  as the {\em Endorsements Ledger}) that acts as 
a decentralized notarization service for the endorsements.
This approach complements the manufacturer signature on an endorsement, 
and it allows a future verifier
in possession of the product to use
the entries in the blockchain -- referred to as {\em endorsement-records} -- 
to validate that:
(1) the endorsement matches the product,
and that (2) the endorsement is source-authentic from the manufacturer 
at the time that the endorsement-record was created (i.e. today)
-- even though at the future time of the verification
the manufacturer
may no longer exists.

The blockchain endorsement-record must also include
pointers to locations (e.g. archival repositories)
on the Internet where copies of the software/firmware
for the product (and the corresponding endorsement objects) may be found. 
Since a manufacturer's endorsement for its software or firmware product
typically include a hash of the software/firmware,
a future verifier who fetches the endorsement-record from the blockchain
obtains assurance that the copy of
the product in its possession is a genuine product
(i.e. unmodified).
Figure~\ref{fig:RIMledger} provides an overview of this idea.

\begin{figure}[t]
\centering
\includegraphics[width=1.0\textwidth, trim={0.0cm 0.0cm 0.0cm 0.0cm}, clip]{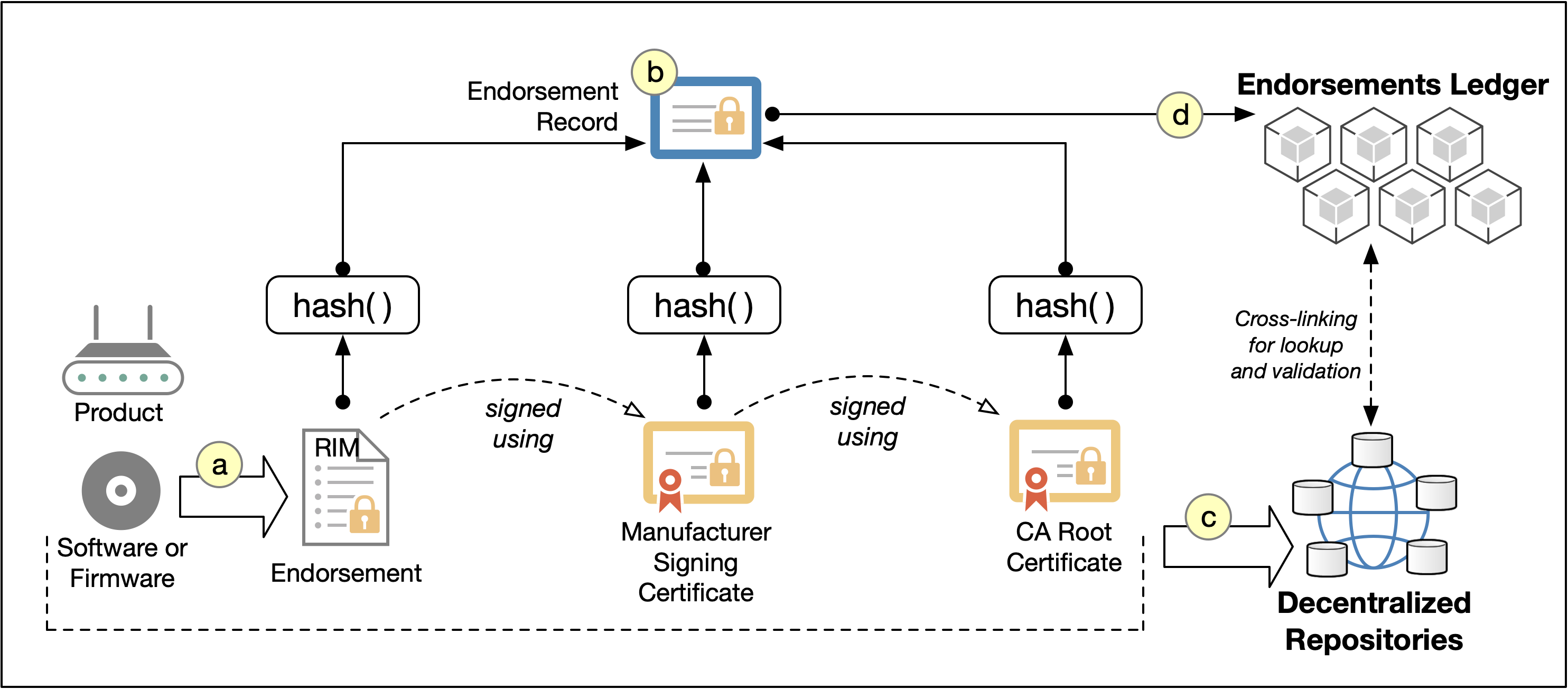}
\caption{Overview of blockchain-based retention of endorsements}
\label{fig:RIMledger}
\end{figure}

In the example of Figure~\ref{fig:RIMledger},
the endorsement-record (Step~(b))
collates the hashes (digests)
of the various objects that support the verification of 
a manufacturer's endorsement
(e.g. the endorsement or RIM~\cite{TCG-IWG-2006-Thomas-Ned-Editors-part2,RFC8520-formatted},
manufacturer's signing certificate,
CA's root certificate, etc.)
from Step~(a).
In essence,
the endorsement-record becomes akin to the root (top) of a Merkle hash-tree.
The manufacturer also stores copies of the relevant verification objects
in a decentralized file repository system (e.g. IPFS~\cite{IPFS})
to ensure the high-availability of these objects (Step~(c)).
The decentralized file repository must ensure availability
of these object long into the future.
Other traditional methods,
such as distributing via CD/DVD discs (e.g. for consumers of the product)
could also be performed.
Finally, the manufacturer transmits the endorsement-record
from Step~(b) onto the Endorsements Ledger in Step~(d).
The endorsement-record (as a transaction on the blockchain) 
needs to include pointers to (e.g. URL/URI)
to the locations of these verification objects,
such that any verifier can later easily fetch these objects
and perform endorsement verification.

\end{itemize}

\section{Conclusions}
\label{sec:Conclusions}

As mentioned in the opening,
we believe that there is a strong role for trusted computing technologies,
and more specially attestations technologies,
in the growing area of blockchain networks.
The security, resiliency and interoperability
of blockchain systems are important factors in the adoption
of blockchain networks as the foundation of the future digital economy.

In the current work we have provided
a high-level review of the notion of attestations,
and described the evolving new standard architecture
for device attestations.
The application of this attestations architecture
to the blockchain environment have been discussed.
The blockchain industry's use of a common standard attestations architecture
can ensure the best chance for the interoperability of systems and networks,
and offers the best path forward towards achieving the survivability of 
various blockchain networks.

Several challenges remain to be addressed as
new modes of implementations of blockchain networks,
such as virtualization and containerization,
become attractive for deployers of blockchains.

~~\\
~~\\



\begin{thebibliography}{10}
\providecommand{\url}[1]{#1}
\csname url@samestyle\endcsname
\providecommand{\newblock}{\relax}
\providecommand{\bibinfo}[2]{#2}
\providecommand{\BIBentrySTDinterwordspacing}{\spaceskip=0pt\relax}
\providecommand{\BIBentryALTinterwordstretchfactor}{4}
\providecommand{\BIBentryALTinterwordspacing}{\spaceskip=\fontdimen2\font plus
\BIBentryALTinterwordstretchfactor\fontdimen3\font minus
  \fontdimen4\font\relax}
\providecommand{\BIBforeignlanguage}[2]{{%
\expandafter\ifx\csname l@#1\endcsname\relax
\typeout{** WARNING: IEEEtran.bst: No hyphenation pattern has been}%
\typeout{** loaded for the language `#1'. Using the pattern for}%
\typeout{** the default language instead.}%
\else
\language=\csname l@#1\endcsname
\fi
#2}}
\providecommand{\BIBdecl}{\relax}
\BIBdecl

\bibitem{Pentland2020a}
\BIBentryALTinterwordspacing
A.~Pentland, ``{Building the New Economy: what we need and how to get there},''
  in \emph{Building the New Economy}, A.~Pentland, A.~Lipton, and T.~Hardjono,
  Eds.\hskip 1em plus 0.5em minus 0.4em\relax {MIT} {P}ress - {Work in Progress
  (WIP)}, 2020. [Online]. Available:
  \url{https://wip.mitpress.mit.edu/new-economy}
\BIBentrySTDinterwordspacing

\bibitem{TPM2003Design}
{Trusted Computing Group}, ``{TPM} {M}ain -- {Part 1} {D}esign {P}rinciples --
  {S}pecification {V}ersion {1.2},'' Trusted Computing Group, {TCG} {P}ublished
  {S}pecification, October 2003, available at
  {http://www.trustedcomputinggroup.org/ resources/ tpm\_main\_specification}.

\bibitem{Proudler2002}
B.~Balacheff, L.~Chen, S.~Pearson, D.~Plaquin, and G.~Proudler, \emph{{T}rusted
  {C}omputing {P}latforms: {TCPA} {T}echnology in {C}ontext}.\hskip 1em plus
  0.5em minus 0.4em\relax New York: Prentice Hall, 2002.

\bibitem{ChallenerYoder2008}
D.~Challener, K.~Yoder, R.~Catherman, D.~Safford, and L.~{Van~Doorn},
  \emph{{Practical Guide to Trusted Computing}}.\hskip 1em plus 0.5em minus
  0.4em\relax New York: IBM Press, 2008.

\bibitem{Proudler2014}
G.~Proudler, L.~Chen, and C.~Dalton, \emph{{T}rusted {C}omputing {P}latforms:
  {TPM2.0} in {C}ontext}.\hskip 1em plus 0.5em minus 0.4em\relax New York:
  Springer, 2014.

\bibitem{HardjonoTPM2008}
T.~Hardjono, ``{B}uilding {T}rust {T}hrough {S}trong {D}igital {I}dentity,''
  \emph{Embedded Computing Design}, pp. 13--18, July 2008.

\bibitem{GervaisKarame2014}
A.~Gervais, G.~O. Karame, V.~Capkun, and S.~Capkun, ``Is bitcoin a
  decentralized currency?'' \emph{{IEEE} Security {\&} Privacy}, vol.~12,
  no.~3, pp. 54--60, 2014.

\bibitem{EyalSirer2014}
I.~Eyal and E.~G. Sirer, ``Majority is not enough: Bitcoin mining is
  vulnerable,'' in \emph{Financial Cryptography and Data Security - 18th
  International Conference, {FC} 2014}, March 2014, pp. 436--454.

\bibitem{Hardjono2018-IEEEGlobalSummit}
T.~Hardjono, ``{Blockchain Interoperability and Survivability},'' September
  2018, presentation 2018 IEEE Global Blockchain Summit, NIST, Gaithersburg, MD
  (17-19 September 2018).

\bibitem{HardjonoLipton2020b}
\BIBentryALTinterwordspacing
T.~Hardjono, A.~Lipton, and A.~Pentland, ``{T}owards an {I}nteroperability
  {A}rchitecture {B}lockchain {A}utonomous {S}ystems,'' 2020, {IEEE}
  {T}ransactions on {E}ngineering {M}anagement (to appear). [Online].
  Available: \url{https://arxiv.org/abs/1805.05934}
\BIBentrySTDinterwordspacing

\bibitem{NIST-80202-2018}
D.~Yaga, P.~Mell, N.~Roby, and K.~Scarfone, ``{B}lockchain {T}echnology
  {O}verview,'' National Institute of Standards and Technology Internal Report
  8202, October 2018, {https://doi.org/10.6028/NIST.IR.8202}.

\bibitem{FATF-Recommendation15-2018}
{FATF}, ``{I}nternational {S}tandards on {C}ombating {M}oney {L}aundering and
  the {F}inancing of {T}errorism and {P}roliferation,'' Financial Action Task
  Force (FATF), {FATF}~{R}evision of {R}ecommendation~{15}, October 2018,
  available at:
  http://www.fatf-gafi.org/publications/fatfrecommendations/documents/fatf-recommendations.html.

\bibitem{Ethereum-POS-2019}
\BIBentryALTinterwordspacing
V.~Buterin, ``{Proof of Stake FAQ},'' August 2019. [Online]. Available:
  \url{https://github.com/ethereum/wiki/wiki/Proof-of-Stake-FAQ}
\BIBentrySTDinterwordspacing

\bibitem{AtzeiBartoletti2016}
\BIBentryALTinterwordspacing
N.~Atzei, M.~Bartoletti, and T.~Cimoli, ``A survey of attacks on ethereum smart
  contracts,'' \emph{{IACR} Cryptology ePrint Archive}, vol. 2016, p. 1007,
  2016. [Online]. Available: \url{http://eprint.iacr.org/2016/1007}
\BIBentrySTDinterwordspacing

\bibitem{McKeen2016}
F.~McKeen, I.~Alexandrovich, I.~Anati, D.~Caspi, S.~Johnson, R.~Leslie-Hurd,
  and C.~Rozas, ``{I}ntel {S}oftware {G}uard {E}xtensions {(Intel~SGX)}
  {S}upport for {D}ynamic {M}emory {M}anagement {I}nside an {E}nclave,'' in
  \emph{Proc. Workshop on Hardware and Architectural Support for Security and
  Privacy {(HASP)} 2016}, Seoul, June 2016,
  http://caslab.csl.yale.edu/workshops/hasp2016/program.html.

\bibitem{Costan2017}
\BIBentryALTinterwordspacing
V.~Costan, I.~Lebedev, and S.~Devadas, \emph{{S}ecure {P}rocessors {P}art~{I}:
  {B}ackground, {T}axonomy for {S}ecure {E}nclaves and {I}ntel {SGX}
  {A}rchitecture}.\hskip 1em plus 0.5em minus 0.4em\relax Boston: Now
  publishers inc., 2017, vol.~11, no. 1-2. [Online]. Available:
  \url{http://dx.doi.org/10.1561/1000000051}
\BIBentrySTDinterwordspacing

\bibitem{Rosenstein-MIT-Athena-1988}
M.~A. Rosenstein, D.~E. Geer, and P.~J. Levine, ``The athena service management
  system,'' in \emph{Proceedings of the {USENIX} Winter Conference. Dallas,
  Texas, USA, January 1988}.\hskip 1em plus 0.5em minus 0.4em\relax {USENIX}
  Association, 1988, pp. 203--211.

\bibitem{SteinerNeuman1988}
J.~G. Steiner, B.~C. Neuman, and J.~I. Schiller, ``Kerberos: An authentication
  service for open network systems,'' in \emph{Proceedings of the {USENIX}
  Winter Conference. Dallas, Texas, USA, January 1988}.\hskip 1em plus 0.5em
  minus 0.4em\relax {USENIX} Association, 1988, pp. 191--202.

\bibitem{Saltzer1974}
J.~H. Saltzer, ``{P}rotection and the {C}ontrol of {I}nformation {S}haring in
  {MULTICS},'' \emph{Communications of the {ACM}}, vol.~17, no.~7, pp.
  388--402, July 1974.

\bibitem{TCG-website}
TCG, ``{ {T}rusted {C}omputing {G}roup },''
  http://www.trustedcomputinggroup.org.

\bibitem{McKeen2013}
F.~Mckeen, I.~Alexandrovich, A.~Berenzon, C.~Rozas, H.~Shafi, V.~Shanbhogue,
  and U.~Savagaonkar, ``{I}nnovative {I}nstructions and {S}oftware {M}odel for
  {I}solated {E}xecution,'' in \emph{Proc. Second Workshop on Hardware and
  Architectural Support for Security and Privacy {HASP2013}}, Tel-Aviv, June
  2013, https://sites.google.com/site/haspworkshop2013/workshop-program.

\bibitem{TCG-IWG-2005-Thomas-Ned-Editors-part1}
\BIBentryALTinterwordspacing
T.~Hardjono and N.~{Smith~(ed)}, ``{TCG} {I}nfrastructure {R}eference
  {A}rchitecture for {I}nteroperability {(Part 1)} -- {S}pecification {V}ersion
  {1.0} {R}ev {1.0},'' Trusted Computing Group, {TCG} {P}ublished
  {S}pecification, June 2005. [Online]. Available:
  \url{https://trustedcomputinggroup.org/wp-content/uploads/IWG\_Architecture\_v1\_0\_r1.pdf}
\BIBentrySTDinterwordspacing

\bibitem{TCG-IWG-2006-Thomas-Ned-Editors-part2}
------, ``{TCG} {I}nfrastructure {W}orking {G}roup architecture {(Part 2)} --
  {Integrity Management} -- {S}pecification {V}ersion {1.0} {R}ev {1.0},''
  Trusted Computing Group, {TCG} {P}ublished {S}pecification, November 2006,
  available at {http://www.trustedcomputinggroup.org/resources}.

\bibitem{RFC3281-formatted}
\BIBentryALTinterwordspacing
S.~Farrell and R.~Housley, ``{A}n {I}nternet {A}ttribute {C}ertificate
  {P}rofile for {A}uthorization,'' April 2002, {IETF}~{S}tandard~{RFC3281}.
  [Online]. Available: \url{http://tools.ietf.org/rfc/rfc3281.txt}
\BIBentrySTDinterwordspacing

\bibitem{FIDO-key-Attestation-2015}
\BIBentryALTinterwordspacing
R.~Lindemann and M.~B. Jones, ``{FIDO 2.0: Key Attestation Format},'' FIDO
  Alliance, FIDO Alliance Proposed Standard, September 2015. [Online].
  Available:
  \url{https://fidoalliance.org/specs/fido-v2.0-ps-20150904/fido-key-attestation-v2.0-ps-20150904.html}
\BIBentrySTDinterwordspacing

\bibitem{GlobalPlatform2012}
\BIBentryALTinterwordspacing
{GlobalPlatform}, ``{GlobalPlatform and the Trusted Computing Group Form Work
  Group to Drive Mobile Security Standards and Solutions},'' June 2012.
  [Online]. Available: \url{https://globalplatform.org}
\BIBentrySTDinterwordspacing

\bibitem{IETF-RATSWG}
\BIBentryALTinterwordspacing
IETF, ``{Remote ATtestation ProcedureS ({RATS}) Working Group -- Approved
  Charter}, {Internet Engineering task Force},'' March 2019. [Online].
  Available: \url{https://datatracker.ietf.org/wg/rats/about/}
\BIBentrySTDinterwordspacing

\bibitem{TCG-ATTEST2020}
\BIBentryALTinterwordspacing
TCG, ``{ Attestations Working Group}, {Trusted Computing Group},'' March 2020.
  [Online]. Available: \url{https://members.trustedcomputinggroup.org}
\BIBentrySTDinterwordspacing

\bibitem{TCG-RIV-2020}
\BIBentryALTinterwordspacing
{TCG}, ``{TCG} {Remote Integrity Verification (RIV): Network Equipment Remote
  Attestation System} {Version 1.0, Rev. 0.9b},'' Trusted Computing Group,
  {TCG} Draft Specifications, June 2019. [Online]. Available:
  \url{https://trustedcomputinggroup.org/wp-content/uploads/TCG-NetEq-Attestation-Workflow-Outline_v1r9b_pubrev.pdf}
\BIBentrySTDinterwordspacing

\bibitem{IETF-rats-network-device-attestation-05}
\BIBentryALTinterwordspacing
G.~Fedorkow, E.~Voit, and J.~{Fitzgerald-McKay}, ``{TPM-based Network Device
  Remote Integrity Verification},'' IETF, Internet-Draft
  {draft-fedorkow-rats-network-device-attestation-05}, April 2020. [Online].
  Available:
  \url{https://datatracker.ietf.org/doc/draft-fedorkow-rats-network-device-attestation/}
\BIBentrySTDinterwordspacing

\bibitem{OpenCompute-website2020}
\BIBentryALTinterwordspacing
{OCP}, ``{Open Compute Project},'' 2020. [Online]. Available:
  \url{https://www.opencompute.org}
\BIBentrySTDinterwordspacing

\bibitem{TCG-Attestations-Arch2020}
N.~{Smith~(ed)}, ``{TCG} {A}ttestation {A}rchitecture,'' Trusted Computing
  Group, {TCG} {D}raft {S}pecification -- {V}ersion~{1.0}, February 2020.

\bibitem{rats-arch-02}
\BIBentryALTinterwordspacing
H.~Birkholz, D.~Thaler, M.~Richardson, N.~Smith, and W.~Pan, ``{Remote
  Attestation Procedures Architecture},'' IETF, Internet-Draft
  {draft-ietf-rats-architecture-02}, March 2020. [Online]. Available:
  \url{https://datatracker.ietf.org/doc/draft-birkholz-rats-architecture/}
\BIBentrySTDinterwordspacing

\bibitem{CokerGuttman2011}
\BIBentryALTinterwordspacing
G.~Coker, J.~Guttman, P.~Loscocco, A.~Herzog, J.~Millen, B.~{O?Hanlon},
  J.~Ramsdell, J.~S. Ariel~Segall, and B.~Sniffen, ``{P}rinciples of {R}emote
  {A}ttestation,'' \emph{International Journal of Information Security},
  vol.~10, pp. 63--81, April 2011. [Online]. Available:
  \url{https://doi.org/10.1007/s10207-011-0124-7}
\BIBentrySTDinterwordspacing

\bibitem{SAMLcore}
{OASIS}, ``{A}ssertions and {P}rotocols for the {OASIS} {S}ecurity {A}ssertion
  {M}arkup {L}anguage ({SAML}) {V2.0},'' March 2005, available on
  {http://docs.oasisopen.org/security/ saml/v2.0/ saml-core-2.0-os.pdf}.

\bibitem{Hardjono2019-IEEEcomms-article}
T.~Hardjono, ``{F}ederated {A}uthorization over {A}ccess to {P}ersonal {D}ata
  for {D}ecentralized {I}dentity {M}anagement,'' \emph{{IEEE} {C}ommunications
  {S}tandards {M}agazine}, vol.~4, no.~3, pp. 32--38, December 2019,
  {A}vailable at {https://arxiv.org/pdf/1906.03552.pdf}.

\bibitem{England2016RIOT}
\BIBentryALTinterwordspacing
P.~England, A.~Marochko, D.~Mattoon, R.~Spiger, S.~Thom, and D.~Wooten, ``{RIoT
  - A Foundation for Trust in the Internet of Things},'' Microsoft Research,
  Tech. Rep. MSR-TR-2016-18, April 2016. [Online]. Available:
  \url{https://www.microsoft.com/en-us/research/publication/riot-a-foundation-for-trust-in-the-internet-of-things/}
\BIBentrySTDinterwordspacing

\bibitem{HardjonoSmith2019c}
\BIBentryALTinterwordspacing
T.~Hardjono and N.~Smith, ``{D}ecentralized {T}rusted {C}omputing {B}ase for
  {B}lockchain {I}nfrastructure {S}ecurity,'' \emph{Frontiers Journal - Special
  Issue on Blockchains}, vol.~2, December 2019. [Online]. Available:
  \url{https://doi.org/10.3389/fbloc.2019.00024}
\BIBentrySTDinterwordspacing

\bibitem{TCG-DICE-Implicit-2018}
\BIBentryALTinterwordspacing
{TCG}, ``{TCG} {Implicit Identity Based Device Attestation} {Version 1.0, Rev.
  0.93},'' Trusted Computing Group, {TCG} Published Specifications, March 2018.
  [Online]. Available:
  \url{https://trustedcomputinggroup.org/wp-content/uploads/TCG-DICE-Arch-Implicit-Identity-Based-Device-Attestation-v1-rev93.pdf}
\BIBentrySTDinterwordspacing

\bibitem{TCG-DICE-Symmetric-2020}
\BIBentryALTinterwordspacing
------, ``{TCG} {Symmetric Identity Based Device Attestation} {Version 1.0,
  Rev. 0.95},'' Trusted Computing Group, {TCG} Published Specifications,
  January 2020. [Online]. Available:
  \url{https://trustedcomputinggroup.org/wp-content/uploads/TCG_DICE_SymIDAttest_v1_r0p95_pub-1.pdf}
\BIBentrySTDinterwordspacing

\bibitem{Kelly-Cerberus-2017}
\BIBentryALTinterwordspacing
B.~Kelly, ``{Project Cerberus Security Architecture Overview Specification},''
  Open Compute Project, Published Specifications, September 2017. [Online].
  Available:
  \url{https://github.com/opencomputeproject/Project_Olympus/blob/master/Project_Cerberus/Project%20Cerberus%20Architecture%20Overview.pdf}
\BIBentrySTDinterwordspacing

\bibitem{R3-website}
{R3CEV}, ``R3,'' 2018, https://www.r3.com.

\bibitem{Miller2018}
\BIBentryALTinterwordspacing
R.~Miller, ``{IBM} teams with {M}aersk on new blockchain shipping solution,''
  \emph{Tech Crunch}, August 2018. [Online]. Available:
  \url{https://techcrunch.com/2018/08/09/ibm-teams-with-maersk-on-new-blockchain-shipping-solution/}
\BIBentrySTDinterwordspacing

\bibitem{Morris2020}
\BIBentryALTinterwordspacing
N.~Morris, ``{12 global pharmaceutical firms join EU blockchain consortium
  PharmaLedger},'' \emph{Ledger Insights}, March 2020. [Online]. Available:
  \url{https://www.ledgerinsights.com/pharmaledger-pharmaceutical-blockchain-eu/}
\BIBentrySTDinterwordspacing

\bibitem{Castillo2020}
\BIBentryALTinterwordspacing
M.~{del~Castillo}, ``{Citi, Goldman Sachs Conduct First Blockchain Equity Swap
  On Ethereum-Inspired Platform},'' \emph{Forbes}, February 2020. [Online].
  Available:
  \url{https://www.forbes.com/sites/michaeldelcastillo/2020/02/06/citi-goldman-sachs-conduct-first-blockchain-equity-swap-on-ethereum-inspired-platform}
\BIBentrySTDinterwordspacing

\bibitem{Bitcoin}
\BIBentryALTinterwordspacing
S.~Nakamoto, ``{Bitcoin: A Peer-to-Peer Electronic Cash System},'' 2008.
  [Online]. Available: \url{https://bitcoin.org/bitcoin.pdf}
\BIBentrySTDinterwordspacing

\bibitem{Buterin2014}
V.~Buterin, ``{Ethereum}: A {N}ext-{G}eneration {C}ryptocurrency and
  {D}ecentralized {A}pplication {P}latform,'' Bitcoin Magazine, Report, January
  2014,
  https://bitcoinmagazine.com/articles/ethereum-next-generation-cryptocurrency-decentralized-application-platform-1390528211/.

\bibitem{StackMachine-Wiki}
{W}ikipedia, ``{Stack Machine},'' 2020, available at
  {https://en.wikipedia.org/wiki/Stack\_machine}, Accessed 7 May 2020.

\bibitem{HardjonoLipton2020a}
T.~Hardjono, A.~Lipton, and A.~Pentland, ``{T}owards a {P}ublic {K}ey
  {M}anagement {F}ramework for {V}irtual {A}ssets and {V}irtual {A}sset
  {S}ervice {P}roviders,'' 2020, {J}ournal of {FinTech} (to appear) --
  Available at {https://arxiv.org/pdf/1909.08607}.

\bibitem{IBM-Blockchain-2019}
\BIBentryALTinterwordspacing
IBM, ``{IBM Blockchain Platform },'' IBM Corporation, Technical Overview,
  September 2019. [Online]. Available:
  \url{https://www.ibm.com/cloud/blockchain-platform}
\BIBentrySTDinterwordspacing

\bibitem{Lardinois2019}
\BIBentryALTinterwordspacing
F.~Lardinois, ``{Microsoft launches a fully managed blockchain service},''
  \emph{Techcrunch}, September 2019. [Online]. Available:
  \url{https://techcrunch.com/2019/05/02/microsoft-launches-a-fully-managed-blockchain-service/}
\BIBentrySTDinterwordspacing

\bibitem{TradecoinSciAm2018}
A.~Lipton and A.~Pentland, ``{B}reaking the {B}ank,'' \emph{Scientific
  American}, vol. 318, no.~1, pp. 26--31, 2018.

\bibitem{IETF-draft-ietf-rats-eat-03}
\BIBentryALTinterwordspacing
G.~Mandyam, L.~Lundblade, M.~Ballesteros, and J.~{O'Donoghue}, ``{The Entity
  Attestation Token (EAT)},'' IETF, Internet-Draft {draft-ietf-rats-eat-03},
  February 2020. [Online]. Available:
  \url{https://datatracker.ietf.org/doc/draft-ietf-rats-eat/}
\BIBentrySTDinterwordspacing

\bibitem{IETF-draft-voit-rats-trusted-path-routing-01}
\BIBentryALTinterwordspacing
E.~Voit, ``{Trusted Path Routing using Remote Attestation},'' IETF,
  Internet-Draft {draft-voit-rats-trusted-path-routing-01}, March 2020.
  [Online]. Available:
  \url{https://datatracker.ietf.org/doc/draft-voit-rats-trusted-path-routing/}
\BIBentrySTDinterwordspacing

\bibitem{NIST-800-193}
\BIBentryALTinterwordspacing
A.~Regenscheid, ``{P}latform {F}irmware {R}esiliency {G}uidelines,'' National
  Institute of Standards and Technology, {NIST} {P}ublication {SP 800-193}, May
  2018. [Online]. Available:
  \url{https://csrc.nist.gov/publications/detail/sp/800-193/final}
\BIBentrySTDinterwordspacing

\bibitem{CSA2018}
\BIBentryALTinterwordspacing
{CSA}, ``{Firmware Integrity in the Cloud Data Center},'' Cloud Security
  Alliance (CSA), Whitepaper, 2018. [Online]. Available:
  \url{https://downloads.cloudsecurityalliance.org/assets/research/firmware/firmware-integrity-in-the-cloud-data-center.pdf}
\BIBentrySTDinterwordspacing

\bibitem{FAAS-Wiki}
{W}ikipedia, ``{Function as A Service (FaaS)},'' 2020, available at
  {https://en.wikipedia.org/wiki/Function\_as\_a\_service}, Accessed 7 May
  2020.

\bibitem{Fowler-Serverless}
M.~Fowler, ``{Serverless Architectures},'' May 2018, available at
  {https://martinfowler.com/articles/serverless.html}, Accessed 7 May 2020.

\bibitem{UTXO-model}
{Bitcoin.org}, ``{Unspent Transaction Output (UTXO)},'' 2020, available at
  {https://bitcoin.org/en/glossary/unspent-transaction-output}, Accessed 7 May
  2020.

\bibitem{HardjonoPentland2016}
T.~Hardjono and A.~Pentland, ``{V}erifiable {A}nonymous {I}dentities and
  {A}ccess {C}ontrol in {P}ermissioned {B}lockchains,'' MIT Connection Science
  \& Engineering, {T}echnical {R}eport, April 2016, available at
  {https://arxiv.org/abs/1903.04584}.

\bibitem{Brickell2004}
E.~Brickell, J.~Camenisch, and L.~Chen, ``{D}irect {A}nonymous {A}ttestation,''
  in \emph{Proceedings of the 11th {ACM} Conference on Computer and
  Communications Security {CCS2004}}.\hskip 1em plus 0.5em minus 0.4em\relax
  ACM, 2004, pp. 132--145.

\bibitem{BrickellLi2012}
E.~Brickell and J.~Li, ``{E}nhanced {P}rivacy {ID}: a {D}irect {A}nonymous
  {A}ttestation {S}cheme with {E}nhanced {R}evocation {C}apabilities,''
  \emph{IEEE Transactions on Dependable and Secure Computing}, vol.~9, no.~3,
  pp. 345--360, 2012.

\bibitem{RFC8520-formatted}
\BIBentryALTinterwordspacing
E.~Lear, R.~Droms, and D.~Romascanu, ``{Manufacturer Usage Description (MUD)
  Specification} {(RFC8520)},'' March 2019. [Online]. Available:
  \url{https://tools.ietf.org/html/rfc8520}
\BIBentrySTDinterwordspacing

\bibitem{IPFS}
{P}rotocol {L}abs, ``{I}nter {P}lanetary {F}ile {S}ystem ({IPFS}),'' 2019,
  available at {https://docs.ipfs.io}, Accessed 23 September 2019.

\end{thebibliography}

\end{document}